\documentclass[pre,aps,twocolumn,superscriptaddress]{revtex4-1}

\usepackage{graphicx}
\usepackage{amssymb,amsfonts,amsmath}
\usepackage{color}
\usepackage{ulem}
\usepackage[hidelinks]{hyperref}
\usepackage[scr=rsfs,cal=boondox]{mathalfa}

\newcommand{\rv}{{\mathbf r}}

\newcommand{\Tr}{{\rm Tr}\,}
\newcommand{\e}{{\rm e}}

\newcommand{\pv}{{\bf p}}

\newcommand{\msphantom}[1]{$\ldots$}

\newcommand{\ms}{\color{black}}

\newcommand{\G}{G}

\newcommand{\eqr}[1]{Eq.~\eqref{#1}}

\newcommand{\mydelete}[1]{{}}

\newcommand{\aas}{{\alpha\alpha'}}
\newcommand{\rmint}{{\rm int}}

\newcommand{\rmexc}{{\rm exc}}
\newcommand{\rmext}{{\rm ext}}

\newcommand{\rmid}{{\rm id}}

\newcommand{\cov}{{\rm cov}}

\newcommand{\go}{{\bar g}}

\newcommand{\rmred}{{\rm red}}
\newcommand{\sbond}{s}

\newcommand{\sdensity}{{\cal s}_\rmexc}
\newcommand{\sumi}{\sum_{i{\ms =1}}^{{\ms N}}}
\newcommand{\Ham}{H_{\ms N}}
\newcommand{\Hams}{H'_{\ms N}}
\newcommand{\Nalpha}{{\cal N}_\alpha}
\newcommand{\Nalphas}{{\cal N}_{\alpha'}}
\newcommand{\calN}{{\cal N}}
\newcommand{\calU}{{\cal U}}

\newcommand{\aalpha}{n}
\newcommand{\alphanew}{a}
\newcommand{\sfG}{{\sf G}}
\newcommand{\nG}{G_n}
\newcommand{\Trmix}{\Tr}
\newcommand{\Trqm}{\Tr}

\begin{document}

\author{Matthias Schmidt}
\affiliation{Theoretische Physik II, Physikalisches Institut, 
  Universit{\"a}t Bayreuth, D-95447 Bayreuth, Germany}
\email{Matthias.Schmidt@uni-bayreuth.de}

\date{26 June 2026, \ms revised version: 21 October 2026}

\title{\ms Entropy functional universality:
  Correlation, response, many-body interaction, 
  multi-component, quantum, hyperfunctional, 
  and Ornstein-Zernike structure}

\begin{abstract}
We give a comprehensive account of the recent {\ms entropy functional
  theory} for the equilibrium statistical mechanics of classical
many-body systems
(\href{https://doi.org/10.48550/arXiv.2606.28240}{arXiv:2606.28240}). The
approach is formally exact and based on a joint grand potential
minimization principle for the one-body density and the global pair
distance distribution. These variational fields depend respectively on
position and on scalar distance, which retains the low computational
complexity of standard density functional theory.  {\ms Correlation
  effects} are contained in a unique excess entropy functional, which
is universal across all systems with pairwise interparticle
potentials.  Functional differentiation yields entropic direct
correlation functionals that generate entropic response and
fluctuation correlation functions via coupled Ornstein-Zernike
equations.  Two alternative proofs are given for the existence and
uniqueness of the underlying metadensity functional map, based on
generalizations of either Levy's constrained search method or
Mermin-Evans proof by contradiction.  Simple excess entropy
approximations yield the standard mean-field and second-virial excess
free energy density functionals.  We describe exact entropic
functional line integrals, make connections to the recent one-body
fluctuation profiles, and generalize the entropy approach beyond
pairwise interparticle potentials {\ms via parametric dependence or
  reduction operators.  We also lay out the formulation for
  multi-component systems, the hyperfunctional for general
  observables, and the quantum entropy functional.}
\end{abstract}

\maketitle

\section{Introduction}
\label{SECintroduction}

The quest to find commonality and to be able to identify universality
across different systems is a guiding principle of physics.
Rationalizing universality can be very powerful even when aiming to
understand and predict very specific phenomena. Besides particular
recent attempts \cite{schmidt2026entropyShort}, an eminently
successful example is the theoretical insight that the electronic
behaviour of molecular systems and of condensed matter in equilibrium
is governed by a universal `intrinsic' density functional that is
independent of the nuclear arrangements \cite{hohenberg1964,
  mermin1965, kohn1999nobel}, when working within the common
Born-Oppenheimer approximation of classical nuclei. Besides kinetic
energy, this intrinsic density functional is generated solely by the
Coulomb and exchange interactions between the electrons.  The
interactions between the nuclei and the electrons are accounted for
separately, via an arguably surprisingly simple explicitly known
bilinear functional.

In contrast to the focus on the Coulombic interparticle interactions,
the {\it classical} version of density functional theory
\cite{evans1979, evans1992, evans2016, hansen2013} is being applied
across a much wider and more varied range of underlying
Hamiltonians. Therein the interparticle potentials describe {\it
  effective} interactions between the classical particles, which can
represent atoms, molecules, or genuine soft matter entities such as
colloids. The application of classical density functional theory to
any concrete system requires one to use (or to formulate or to train)
an intrinsic free energy density functional that is specific to the
particular interparticle interaction potential of the targeted model.
Two of the most widely used analytical approximate free energy density
functionals are Rosenfeld's fundamental-measure theory for hard
spheres \cite{rosenfeld1989}, together with its modified
descendants~\cite{roth2010}, and the mean-field (or `random-phase')
approximation, which is bilinear in the density profile with the pair
potential providing the coupling; see also recent interesting work
\cite{belloni2026} that revisits the weighted-density
approximation~\cite{denton1989, denton1990, denton1991}.

Classical density functional theory receives a significant current
boost through the systematic incorporation of machine learning
methods, which can be used in flexible ways to give accurate and
computationally efficient functional representations that encapsulate
the many-body problem. Typical examples of neural functional work has
addressed the physics of the iconic fluids of hard spheres
\cite{sammueller2023neural} and hard rods \cite{sammueller2023neural,
  sammueller2023whyNeural}, of pure \cite{sammueller2023neural,
  sammueller2024attraction} and binary~\cite{robitschko2025mixShort,
  robitschko2026mixLong, zhou2026azeoptropic} Lennard-Jones systems,
as well as of charged systems~\cite{bui2024neuralrpm,
  bui2025electromechanics, bui2025dielectrocapillarity} and systems
interacting with {\it ab initio} interparticle interaction potentials
\cite{bui2026abinitio}.  The neural functional method also stimulated
work to formulate significant extensions of standard classical density
functional concepts, such as the metadensity functional formulation
\cite{kampa2024meta, kampa2026pairmatching, kampa2026spherical} for
capturing explicitly the functional dependence on the pair potential,
and the hyperdensity functional framework
\cite{sammueller2024hyperDFT, sammueller2024whyhyperDFT,
  sammueller2024multihyperDFT} for the behaviour of general order
parameters and observables of interest. In nonequilibrium the relevant
exact power functional maps \cite{schmidt2022rmp} have been shown to
be accurately accessible \cite{delasheras2023perspective,
  zimmermann2024ml}.  The local learning method by Samm\"uller {\it et
  al.}~\cite{sammueller2023neural} provides an efficient means to
access the respective functional dependencies, and detailed
quantitative comparisons with simulation reference data were carried
out \cite{dijkman2024ml, sammueller2024pairmatching,
  kampa2026pairmatching, kampa2026spherical}.

Here we give an in-depth {\ms account} of the entropy functional
theory of Ref.~\cite{schmidt2026entropyShort}. The key concept of the
approach is to treat the external and interparticle potential energies
on an equal footing. {\ms Technically this is achieved by
  supplementing the standard density operator~$\hat\rho(\rv)$,
  where~$\rv$ denotes (vectorial) position, by a `pair reduction
  operator' $\hat G(r)$, where $r$ denotes (scalar) interparticle pair
  distance. This setup avoids having to deal with the full
  inhomogeneous two-body density distribution, as is implied in
  previous discussions of entropy in the classical density functional
  context~\cite{percus1996, percus1994aspects, percus1989entropy}.
  Our setup provides a full formal account of entropy and hence it is
  distinct from mere truncation of $n$-body factorization at the pair
  level; see Ref.~\cite{zu2020entropy} for a discussion thereof in a
  modern context.  We demonstrate that the reduction operator concept
  continues to apply beyond pairwise interparticle forces and that it
  leads to efficient scaling of computational demand with
  increasing~$n$.}

For recent insightful work that addresses entropy, we refer the Reader
to the studies by Frusawa \cite{frusawa2018, frusawa2019,
  frusawa2021}, Keffer and coworkers \cite{nicholson2021, sluss2022},
and Shirai {\it et~al.}~\cite{shirai2025}.  A specific form of local
entropy density was introduced in
Ref.~\cite{schmidt2011internalEnergy} and this was shown to be similar
to the local thermal susceptibility which constitutes a local measure
of entropic fluctuations \cite{eckert2020, eckert2023fluctuation,
  coe2022pre}.  The recent approach~\cite{schmidt2026entropyShort}
identifies the entropy functional~$S[\rho,G]$ as the central and
universal object. Its existence and uniqueness have a broad range of
theoretical implications and consequences, as we develop in detail in
the present contribution.

\smallskip

The manuscript is organized as follows. We lay out the statistical
mechanics in Sec.~\ref{SECstatisticalMechanics}, starting with the
specification of the microscopic many-body Hamiltonian and the
relevant density and distance observables in Sec.~\ref{SECmodel}.  To
provide some background, in Sec.~\ref{SECcdftOverview} the main
features of standard classical density functional theory are
summarized.  We give then in Sec.~\ref{SECoverview} an abridged
overview of the key elements of the entropy functional framework
\cite{schmidt2026entropyShort}.  Section~\ref{SECidealFunctionals}
contains derivations of the standard ideal entropy and kinetic energy
density functionals.  

The entropy functional framework is developed in
Sec.~\ref{SECentropyFunctionalFramework}, beginning with details on
the entropic constrained search method that is used for the formal
construction of the excess entropy functional in
Sec.~\ref{SECentropyAppendixOmegaDerivation}.  In
Sec.~\ref{SECrelationshipWithStandardDFT} the relationship with
standard classical density functional theory is laid out via the
method of partial functional minimization. In
Section~\ref{SECchangeOfVariables} we introduce a scaled and hence
intensive global distance distribution $\go(r)$ and lay out the
details for the change of variables from $G(r)$ to $\go(r)$.  In
Section~\ref{SECentropyFunctionalIntegration} the relevant entropic
functional line integrals are described.

Entropic response and correlation formalism are presented in
Sec.~\ref{SECresponseAndCorrelation}, starting in
Sec.~\ref{SECsecondOrderResponseFunction} with second-order entropic
response and fluctuation correlation functions that follow from
functional differentiation.  {\ms Section~\ref{SECsecondOrderDirect}
  describes second-order entropic direct correlation functionals,
  which are obtained as functional derivatives of the excess entropy
  functional.}  In Section~\ref{SECOrnsteinZernikeExcessEntropy} we
present a set of coupled exact entropic Ornstein-Zernike equations
that relate these correlation functions to each other.  In
Sec.~\ref{SECentropyFluctuationProfiles} the relationship of the
entropy functional theory with local fluctuation profiles is shown.

Various specific aspects of the entropy functional are described in
Sec.~\ref{SECentropyFunctionalForms}.  We start in
Sec.~\ref{SECentropyLocalizedEntrropyFunctional} by presenting
spatially localized version of the excess entropy functional.
Section~\ref{SECreducedFreeEnergyPicture} lays out a reduced free
energy constrained search formulation that is equivalent to the
entropic formulation.  Section~\ref{SECpairEntropyApproximation}
presents a pairwise excess entropy approximation that we demonstrate
to be analogous to the second-virial excess free energy approximation.
In Sec.~\ref{SECmeanField} we show how a simple entropy approximation
reduces to the standard mean-field free energy functional
approximation.

{\ms Several generalizations of the framework are presented in
  Sec.~\ref{SECgeneralization}.  The incorporation of many-body
  interparticle interactions into the entropy framework is laid out
  via parametric dependence in
  Sec.~\ref{SECentropyManyBodyInteractions}. The universal entropy
  functional resolution for the many-body interacting case is obtained
  via the reduction operator formalism presented in
  Sec.~\ref{SECviaReductionOperators}.  The computational benefits of
  this method are discussed in
  Sec~\ref{SECmanyBodyPracticalities}. The entropic hyperfunctional
  theory for the thermal behaviour of general observables is presented
  in Sec.~\ref{SEChyperfunctionalEntropy}.  Classical multi-component
  systems are addressed in Sec.~\ref{SECmultiComponent}. The quantum
  version of the entropy functional theory is formulated in
  Sec.~\ref{SECquantumEntropy}.}  The penultimate
Sec.~\ref{SECentropyProofByContradiction} demonstrates the relevant
functional map via proof by contradiction.  In
Sec.~\ref{SECconclusions} we give our conclusions.

\section{Statistical mechanics}
\label{SECstatisticalMechanics}

We start in Sec~\ref{SECmodel} with a description of the many-body
Hamiltonian and define the density and distance observables,
$\hat\rho(\rv)$ and $\hat G(r)$, that are relevant for the subsequent
entropy functional construction.  The essentials of standard classical
density functional theory are summarized in Sec.~\ref{SECcdftOverview}
to provide a backdrop for the present method.  We give an abridged
overview of the key elements of the entropy functional
framework~\cite{schmidt2026entropyShort} in Sec.~\ref{SECoverview}.
To demonstrate some of the involved concepts for a simplistic standard
case, the ideal entropy and kinetic energy density functionals are
derived in Sec.~\ref{SECidealFunctionals}.

\subsection{Microscopic many-body model}
\label{SECmodel}
The standard Hamiltonian $\Ham$ {\ms consists of a sum of kinetic
  energy, interparticle interaction potential, and external potential
  energy. This} is expressed as
\begin{align}
  \Ham &= \sumi \frac{\pv_i^2}{2m} + u(\rv^N) + \sumi V_\rmext(\rv_i),
  \label{EQentropyHamiltonianAPX}
\end{align}
where $\pv_i$ is the momentum of particle $i$, the particle mass is
denoted by $m$, the interparticle potential $u(\rv^N)$ depends on the
position coordinates $\rv^N=\rv_1,\ldots, \rv_N$ of all $N$ particles,
and $V_\rmext(\rv)$ is the extenal potential, expressed here as a
function of the generic position variable~$\rv$. {\ms In keeping with
  standard notation \cite{evans1979}, we have made the dependence of
  the Hamiltonian \eqref{EQentropyHamiltonianAPX} on the total number
  of particles $N$ explicit via the subscript of~$\Ham$. The particles
  are considered to be indistinguishable, such that $u(\rv^N)$ is
  invariant under particle index permutations, and general forms of
  $u(\rv^N)$ will be considered below in
  Secs.~\ref{SECentropyManyBodyInteractions}--\ref{SECmanyBodyPracticalities}.}

{\ms Here we start with interparticle interaction potentials
  $u(\rv^N)$ that consist of merely pairwise contributions. This
  allows one to express the
  Hamiltonian~\eqref{EQentropyHamiltonianAPX} in the following
  re-written form:}
\begin{align}
  \Ham &= \sumi \frac{\pv_i^2}{2m} + \int_0^\infty dr \hat G(r)\phi(r)
  + \int d\rv \hat\rho(\rv) V_\rmext(\rv),
  \label{EQentropyHamiltonianRewrittenAPX}
\end{align}
{\ms where in an important class of systems the particles} interact
solely via a pair potential $\phi(r)$, where $r$ denotes the
interparticle pair distance. {\ms Then the interparticle interaction
  potential $u(\rv^N)$, as it appears in
  \eqr{EQentropyHamiltonianAPX},} can be expressed as
\begin{align}
  u(\rv^N) &= \frac{1}{2}\sum_{i=1}^N \sum_{j=1,j\neq i}^N\phi(|\rv_i-\rv_j|)
  \label{EQentropyInterparticlePotentialPairwiseAPX}
  \\
  &=  \int_0^\infty dr \hat G(r) \phi(r),
  \label{EQentropyInterparticlePotentialReWritingAPX}
\end{align}
{\ms where the standard form
  \eqref{EQentropyInterparticlePotentialPairwiseAPX} involves the sum
  over all distinct pairs $i,j$ with correction for double
  counting. This procedure is re-expressed in the re-written
  form~\eqref{EQentropyInterparticlePotentialReWritingAPX} via} $\hat
G(r)$ as the global pair reduction operator (phase space function),
which is defined as \cite{kampa2024meta, kampa2026pairmatching,
  kampa2026spherical}:
\begin{align}
  \hat G(r) &=
  \frac{1}{2}\sum_{i=1}^N \sum_{j=1,j\neq i}^N \delta(r-|\rv_i-\rv_j|).
  \label{EQentropyGhatAPX}
\end{align}
{\ms The operator $\hat G(r)$ can be viewed as an `instantaneous'
  global histogram of the interparticle pair distances across the
  entire system. We defer the generalization to $n$-body reduction
  operators $\hat G_n(R^N)$ that are relevant in the presence of
  many-body interparticle interaction potentials to
  Secs.~\ref{SECviaReductionOperators} and
  \ref{SECmanyBodyPracticalities}.}  In a formally analogous way to
\eqr{EQentropyInterparticlePotentialReWritingAPX}, the external
potential energy [last term in Eqs.~\eqref{EQentropyHamiltonianAPX}
  and \eqref{EQentropyHamiltonianRewrittenAPX}] is re-written in the
standard way as
\begin{align}
  \sumi V_\rmext(\rv_i) &= \int d\rv \hat\rho(\rv) V_\rmext(\rv),
  \label{EQentropyVextReWritingAPX}
\end{align}
where the density operator has the form \cite{evans1979, hansen2013}:
\begin{align}
  \hat\rho(\rv) &= \sumi \delta(\rv-\rv_i).
  \label{EQdensityOperatorAPX}
\end{align}

{\ms Being able to identify the operator form
  \eqref{EQentropyVextReWritingAPX} of the global external potential
  energy is an important standard step in the construction of free
  energy-based density functional theory.  Here we supplement this
  mathematical structure by expressing the pair interaction energy in
  the pair distance operator form \eqref{EQentropyGhatAPX}.} The thus
re-written Hamiltonian \eqref{EQentropyHamiltonianRewrittenAPX}
expresses the interparticle and external potential energies on an
equivalent formal footing and this will be the basis for building the
entropy functional framework.

\subsection{Classical density functional theory}
\label{SECcdftOverview}

We first provide some context and hence give an abridged account of
several key results of standard classical density functional theory
\cite{evans1979, evans1992, evans2016, hansen2013}. The grand
potential density functional $\Omega[\rho]$ consists of a sum of
intrinsic, external and chemical contributions, according to
\begin{align}
  \Omega[\rho] &= F[\rho] + \int d\rv \rho(\rv)[V_\rmext(\rv)-\mu],
  \label{EQcdftOmegaFunctional}
\end{align}
where $\mu$ denotes the chemical potential. The intrinsic Helmholtz
free energy density functional~$F[\rho]$ depends on temperature $T$
and on the form of the interparticle potential $u(\rv^N)$ and these
dependencies are suppressed in the (standard) notation. Crucially,
$F[\rho]$ is independent both of the form of the external potential
$V_\rmext(\rv)$ and of the chemical potential $\mu$, which both
feature merely via linear dependance in the position integral in
\eqr{EQcdftOmegaFunctional}. The existence and uniqueness of the
density functional \eqref{EQcdftOmegaFunctional} can be proven via the
Mermin-Evans proof argument by contradiction \cite{mermin1965,
  evans1979, hansen2013} or by Levy's constrained search method
\cite{levy1979, dwandaru2011}.

The grand potential density functional \eqref{EQcdftOmegaFunctional}
is minimized by the `true' equilibrium density profile $\rho_0(\rv)$,
such that
\begin{align}
  \frac{\delta\Omega[\rho]}{\delta\rho(\rv)} \Big|_{\rho=\rho_0} &= 
  0 \qquad {\rm (min)},
\end{align}
where the notation indicates that the equilibrium density profile
$\rho_0(\rv)$ is inserted after the density functional derivative
$\delta/\delta\rho(\rv)$ has been taken. Then evaluating the grand
potential density functional \eqref{EQcdftOmegaFunctional} at the
minimizer $\rho_0(\rv)$ gives 
\begin{align}
  \Omega[\rho_0] &= \Omega_0,
  \label{EQcdftOmegaValue}
\end{align}
where $\Omega_0$ is the value of the grand potential; the subscripts 0
will be dropped for brevity of notation.  The minimization condition
\eqref{EQcdftOmegaFunctional} can be cast into the following more
explicit (Euler-Lagrange) form:
\begin{align}
  c_1(\rv;[\rho]) &= \ln\big(\rho(\rv)\Lambda^d\big) 
  + \beta V_\rmext(\rv) - \beta\mu,
  \label{EQcdftEulerLagrange}
\end{align}
where $\Lambda$ is the thermal de Broglie wavelength
\cite{hansen2013}, $d$ denotes the spatial dimensionality,
$\beta=1/(k_BT)$ is inverse temperature, with $k_B$ denoting the
Boltzmann constant, and $T$ absolute temperature. The one-body direct
correlation functional $c_1(\rv;[\rho])$ is defined as the following
functional derivative:
\begin{align}
  c_1(\rv;[\rho]) &= 
  -\frac{\delta \beta F_\rmexc[\rho]}{\delta\rho(\rv)},
  \label{EQcdftOneBodyDirectCorrelationFunctional}
\end{align}
where the excess free energy functional $F_\rmexc[\rho]$ constitutes
the nontrivial (over ideal gas) contribution to the intrinsic
Helmholtz free energy functional $F[\rho]$; details are provided
below.  Once $c_1(\rv;[\rho])$ is known, typically from using an
approximation scheme or via neural functional learning
\cite{sammueller2023neural, sammueller2023whyNeural,
  sammueller2024attraction}, then the Euler-Lagrange
equation~\eqref{EQcdftEulerLagrange} constitutes an implicit
functional relation that determines the equilibrium density
profile. All thermodynamical properties of the system then follow via
the grand potential relationship \eqref{EQcdftOmegaValue}. Two- and
higher-body correlation functions can be accessed systematically by
the test particle \cite{percus1962}, Ornstein-Zernike
\cite{hansen2013}, and metadensity \cite{evans1992,
  kampa2026pairmatching} routes.

\subsection{Entropy functional overview}
\label{SECoverview}

We describe several key elements of the entropic density functional
approach of Ref.~\cite{schmidt2026entropyShort}. Similar to the
construction of standard density functional theory~\cite{evans1979,
  evans1992, evans2016, hansen2013, dwandaru2011}, the starting point
is provided by Mermin's functional,
\begin{align}
  \beta \Omega_M[f] &= \Tr f (\ln f + \beta \Ham - \beta \mu N),
  \label{EQentropyMerminFunctionalDefinitionAPX}
\end{align}
where $\Tr\,\cdot\, = \sum_{N=0}^\infty (h^{dN}N!)^{-1} \int d\rv^N
d\pv^N\,\cdot\,$ indicates the standard classical grand canonical
`trace' integral operator and $h$ the Planck constant. The $N$-body
probability distribution function $f(\rv^N, \pv^N)$ depends on the
classical phase space variables $\rv^N, \pv^N$, with
$\pv^N=\pv_1,\ldots,\pv^N$ denoting all momenta. The many-body
distribution function $f$ constitutes the `trial' functional argument
in Mermin's functional \eqref{EQentropyMerminFunctionalDefinitionAPX}
with the mere requirements of being non-negative, $f\geq 0$, and
normalized, $\Tr f=1$, but otherwise unspecified.

To construct the metadensity functional dependence, a constrained
search is performed over the function space of many-body distribution
functions \cite{levy1979, dwandaru2011, schmidt2026entropyShort}:
\begin{align}
  \Omega[\rho,G] &= \min_{f\to\rho,G} \Omega_M[f].
  \label{EQentropyOmegaFunctionalFromMerminAPX}
\end{align}
The arrow notation, $f\to\rho,G$, indicates that the
minimization~\eqref{EQentropyOmegaFunctionalFromMerminAPX} is
performed under the following {\it two} simultaneous constraints:
\begin{align}
  \rho(\rv) &= \Tr f \hat\rho(\rv),
  \label{EQentropyConstraint1APX}\\
  G(r) &= \Tr f \hat G(r).
  \label{EQentropyConstraint2APX}
\end{align}
{\ms The trial fields \eqref{EQentropyConstraint1APX} and
  \eqref{EQentropyConstraint2APX} constitute respectively the standard
  one-body density `profile' $\rho(\rv)$ and the mean (global) pair
  reduction function $G(r)$; we recall the
  definition~\eqref{EQentropyGhatAPX} of the pair reduction
  operator~$\hat G(r)$ together with the microscopic form
  \eqref{EQdensityOperatorAPX} of the density
  operator~$\hat\rho(\rv)$.}

It can then be shown (we give details below) that the resulting grand
potential metadensity functional
\eqref{EQentropyOmegaFunctionalFromMerminAPX} splits into the
following sum:
\begin{align}
  \Omega[\rho,G] &=  K[\rho] - TS[\rho,G] 
  + \int_0^\infty dr G(r) \phi(r)
  \notag\\&\quad 
  +\int d\rv  \rho(\rv) [ V_\rmext(\rv) -  \mu],
  \label{EQentropyOmegaDecompositionAPX}
\end{align}
where $K[\rho]$ is the kinetic energy density functional [the explicit
  form \eqref{EQentropyKineticEnergyFunctional} of the classical
  version is given in Sec.~\ref{SECidealFunctionals} below], $S[\rho,
  G]$ is the entropy metadensity functional, where the prefix
terminology refers to non-trivial functional dependence beyond the
density profile, here on $G(r)$ rather than on the pair potential
$\phi(r)$ \cite{kampa2024meta, kampa2026pairmatching,
  kampa2026spherical}. The third and fourth terms in the decomposition
\eqref{EQentropyOmegaDecompositionAPX} comprise respectively the
interparticle potential energy and the sum of the external potential
energy and chemical potential contributions. The nontrivial part in
the grand potential functional~\eqref{EQentropyOmegaDecompositionAPX}
is~$S[\rho,G]$. This is given formally by the following constrained
search entropy maximization:
\begin{align}
  & S[\rho,G] = 
  \max_{f\to \rho,G}
  \Tr f\Big( -k_B \ln f 
  -\frac{1}{T}\sumi\frac{\pv_i^2}{2m}
  \Big) + \frac{K[\rho]}{T},
  \label{EQentropyViaConstrainedMaximizationAPX}
\end{align}
{\ms 
and we provide details below together with the purely configurational
alternative
\eqref{EQentropyViaConstrainedMaximizationConfigurational}.

The very existence of the specific form of the constrained search
functional \eqref{EQentropyViaConstrainedMaximizationAPX} is
noteworthy, as this generalizes the standard free energy functional of
classical density functional theory \cite{evans1979, evans1992,
  evans2016, hansen2013, levy1979, dwandaru2011}. We recall that the
Levy search involves `scanning' over the function space of phase space
distribution functions $f(\rv^N, \pv^N)$~\cite{levy1979,
  dwandaru2011}. Thereby the target is to identify the specific form
of $f(\rv^N, \pv^N)$ that maximizes the standard many-body form of
entropy, see the argument of the constrained maximization in
\eqr{EQentropyViaConstrainedMaximizationAPX}.  The additive many-body
kinetic term ensures that the momentum degrees of freedom are taken
into account correctly. The resulting kinetic energy offset is removed
by the last term, $K[\rho]/T$.  The emergent functional dependence of
the entropy is then on the constraint fields $\rho(\rv)$ and $G(r)$.

The above procedure implies a very significant reduction of argument
complexity, due to the respective dependence of $\rho(\rv)$ and $G(r)$
on a mere single position variable $\rv$ and on scalar pair
distance~$r$. Note the striking contrast to the $N$-body phase space
dependence of $f(\rv^N, \pv^N)$ in Mermin's functional
\eqref{EQentropyMerminFunctionalDefinitionAPX}.
However the constrained search method does not by itself deliver an
{\it explicit} form of the dependence of $S[\rho,G]$ on its functional
arguments.
One cannot hope to find this in closed analytical exact form, as this
amounts to the exact solution of the thermal equilibrium physics of
the full many-body problem in general inhomogeneous conditions.
Nevertheless the status of the entropy
functional~\eqref{EQentropyViaConstrainedMaximizationAPX} is an
important one, as this provides a systematic starting point for
developing the entirety of the equilibrium statistical mechanics. The
motivation behind developing the formal structure is to guide the
construction of systematic machine learning schemes as well as
fundamental analytical approximations, as we motivate in more detail
further below.

}

Working on the level of resolution that is provided by the joint set
of fields $\rho(\rv)$ and $G(r)$ allows one to express explicitly the
thermal mean of the total energy $E[\rho,G]$ in the form of the
following sum of the kinetic and the two potential energy
contributions:
\begin{align}
  E[\rho,G] &= K[\rho] 
  + \int_0^\infty dr G(r) \phi(r)
  + \int d\rv \rho(\rv) V_\rmext(\rv).
  \label{EQmeanEnergyFunctionalAPX}
\end{align}
The energy sum \eqref{EQmeanEnergyFunctionalAPX} hence comprises the
first, third and (partial) fourth terms in the grand potential
functional~\eqref{EQentropyOmegaDecompositionAPX}, and we turn to the
remaining contributions in the following.

It is useful to first define the trivial mean particle number density
functional $N[\rho]$. This is given simply as the position integral
over the density profile,
\begin{align}
  N[\rho] &= \int d\rv \rho(\rv),
  \label{EQmeanParticleNumberDensityFunctional}
\end{align}
as follows from spatial integration over the density constraint
\eqref{EQentropyConstraint1APX} and recognizing that $\int d\rv
\hat\rho(\rv)=N$, see the definition \eqref{EQdensityOperatorAPX} of
the density operator $\hat\rho(\rv)$.

The definitions \eqref{EQmeanEnergyFunctionalAPX} and
\eqref{EQmeanParticleNumberDensityFunctional} allow one to express the
grand potential metadensity functional
\eqref{EQentropyOmegaDecompositionAPX} in the following split form:
\begin{align}
  \Omega[\rho,G] &=
  E[\rho,G]  -TS[\rho,G]  - \mu N[\rho],
  \label{EQmetaentropyOmegaViaEnergyEntropyAPX}
\end{align}
which constitutes a functional version \cite{schmidt2026entropyShort}
of the standard thermodynamic decomposition of the grand potential
into energetic, entropic and particle number contributions
\cite{hansen2013}.

The equilibrium grand potential $\Omega_0 = -k_BT\ln\Xi$, where $\Xi =
\Tr \e^{-\beta (\Ham-\mu N)}$ is the grand canonical partition sum, then
follows as the value of the metadensity functional
\eqref{EQmetaentropyOmegaViaEnergyEntropyAPX} at its minimum:
\begin{align}
  \Omega_0 = \min_{\rho,G} \Omega[\rho,G].
  \label{EQentropyMinimizationPrincipleAPX}
\end{align}
The minimization principle \eqref{EQentropyMinimizationPrincipleAPX}
implies that the following two Euler-Lagrange equations hold at the
minimum,
\begin{align}
  \frac{\delta \Omega[\rho,G]}{\delta\rho(\rv)}\Big|_G &= 0
  \quad\text{(min)},
  \label{EQentropyEL1rawAPX}  \\
  \frac{\delta \Omega[\rho,G]}{\delta G(r)}\Big|_\rho &=0
  \quad\text{(min)}.
  \label{EQentropyEL2rawAPX}
\end{align}
The minimizers are the equilibrium density profile $\rho_0(\rv)$ and
global pair distance distribution function $G_0(\rv)$, as obtained
from evaluating the constraints \eqref{EQentropyConstraint1APX} and
\eqref{EQentropyConstraint2APX} at the global minimum of Mermin's
functional \eqref{EQentropyMerminFunctionalDefinitionAPX}. Recall that
Mermin's functional $\Omega_M[f]$ operates on phase space distribution
functions $f$ and the minimizer is the equilibrium distribution
function $f_0(\rv^N, \pv^N)= \e^{-\beta(\Ham-\mu N)}/\Xi$. For
completeness, corresponding equilibrium averages are then obtained by
$\langle \, \cdot \, \rangle = \Tr \,\cdot\, f_0$.  Evaluating the
metadensity functional \eqref{EQmetaentropyOmegaViaEnergyEntropyAPX}
at the minimizers yields the equilibrium grand potential,
\begin{align}
  \Omega_0 = \Omega[\rho_0,G_0],
  \label{EQOmegaFromEvaluationAtTheMinimizers}
\end{align}
and we leave away the subscript zero of $\rho_0(\rv)$
and $G_0(\rv)$ for brevity of notation in the following.

To identify the nontrivial correlation contribution to the grand
potential functional \eqref{EQmetaentropyOmegaViaEnergyEntropyAPX}, we
split the total entropy functional $S[\rho,G]$ into separate ideal gas
and excess (over ideal gas) contributions,
\begin{align}
  S[\rho,G] &= S_\rmid[\rho] + S_\rmexc[\rho,G],
  \label{EQentropySidSexcSplittingAPX}
\end{align}
where the ideal gas entropy 
functional $S_\rmid[\rho]$ is known analytically and given below [see
  \eqr{EQentropySidOneDefinition} in Sec.~\ref{SECidealFunctionals}].
The excess entropy metadensity functional $S_\rmexc[\rho,G]$ is unique
and universal, such that its knowledge allows one, in principle, to
treat any (admissible) form of the pair potential~$\phi(r)$. We recall
that the pair potential features only via its explicit occurrence in
the bilinear interparticle potential energy functional, cf.\ the
second term in the total energy functional
\eqref{EQmeanEnergyFunctionalAPX}.

It is natural to define {\it entropic} direct correlation functionals
via the following first functional derivatives of the excess entropy
metadensity functional:
\begin{align}
  c_\rho(\rv;[\rho,G])  &= 
  \frac{\delta S_\rmexc[\rho,G]/k_B}{\delta\rho(\rv)}\Big|_G,
  \label{EQentropyDirectEntropic1APX}\\
  c_G(r;[\rho,G]) &= 
  \frac{\delta S_\rmexc[\rho,G]/k_B}{\delta G(r)}\Big|_\rho.
  \label{EQentropyDirectEntropic2APX}
\end{align}
The notation in Eqs.~\eqref{EQentropyDirectEntropic1APX} and
\eqref{EQentropyDirectEntropic2APX} indicates that the functional
derivatives are performed upon keeping the second respective
functional argument constant. Hence the functional derivatives can be
considered as being partial, which is standard.

Using the definitions \eqref{EQentropyDirectEntropic1APX} and
\eqref{EQentropyDirectEntropic2APX} of the entropic direct correlation
functionals allows one to express the Euler-Lagrange equations
\eqref{EQentropyEL1rawAPX} and \eqref{EQentropyEL2rawAPX} in the form
of the following two coupled functional self-consistency equations:
\begin{align}
  c_\rho(\rv;[\rho,G])  &= 
  \ln\big(\rho(\rv)\Lambda^d\big) + \beta V_\rmext(\rv) - \beta\mu,
  \label{EQentropyEL1rhoGAPX}\\
  c_G(r;[\rho,G]) &= \beta\phi(r).
  \label{EQentropyEL2rhoGAPX}
\end{align}
Solving the system of equations \eqref{EQentropyEL1rhoGAPX} and
\eqref{EQentropyEL2rhoGAPX} allows one, in principle, to determine the
equilibrium density profile~$\rho(\rv)$ as well as the (global) pair
reduction function~$G(r)$ for given forms of the external potential
$V_\rmext(\rv)$ and the pair potential $\phi(r)$, and for fixed
thermodynamic parameters $\beta$ and $\mu$. In practice, this requires
one to have a handle on the two entropic direct correlation
functionals on the left hand sides of Eqs.~\eqref{EQentropyEL1rhoGAPX}
and \eqref{EQentropyEL2rhoGAPX} or, equivalently, on their functional
generator $S_\rmexc[\rho,G]$; we recall the functional derivative
structure \eqref{EQentropyDirectEntropic1APX} and
\eqref{EQentropyDirectEntropic2APX}.

Once the equilibrium solutions for $\rho(\rv)$ and $G(r)$ are
obtained, their insertion into the grand potential metadensity
functional \eqref{EQOmegaFromEvaluationAtTheMinimizers} gives the
equilibrium grand potential $\Omega_0$, which determines formally the
thermodynamics.

\subsection{Classical ideal gas functionals}
\label{SECidealFunctionals}

It is useful to describe the ideal gas physics from the current
entropy angle. We recall that the density functionals for kinetic
energy, $K[\rho]$, and for ideal entropy, $S_\rmid[\rho]$, were used
in Sec.~\ref{SECoverview} to identify the nontrivial excess entropy
metadensity functional $S_\rmexc[\rho,G]$; see the respective
occurrences in the mean potential
energy~\eqref{EQmeanEnergyFunctionalAPX} and in the entropic
ideal-excess splitting \eqref{EQentropySidSexcSplittingAPX}.  We first
describe general mechanisms that are relevant in the present context,
{\ms before laying out the explicit calculation for the ideal gas at
  the end of the subsection.}

We use the fundamental thermodynamical relationship of entropy~$S$
with the grand potential,
\begin{align}
  S = -\frac{\partial \Omega_0}{\partial T}.
  \label{EQentropyFromOmega0}
\end{align}
{\ms 
The partial temperature derivative \eqref{EQentropyFromOmega0} is
built in the standard way upon fixing all quantities that determine
the physical system: these are the chemical potential $\mu$, the
system volume, as well as the forms of the external potential
$V_\rmext(\rv)$ and interparticle interaction potential~$u(\rv^N)$.
The derivative \eqref{EQentropyFromOmega0} will, in general, result in
differential changes to the density profile $\rho(\rv)$ and to the
global pair reduction function $G(r)$, which hence both vary in the
physically correct way of the temperature response that a real system
displays \cite{eckert2020, eckert2023fluctuation}.}

It is useful to demonstrate that the parametric temperature derivative
\eqref{EQentropyFromOmega0} is consistent with the statistical
mechanical underpinnings, as follows via standard elementary
calculation: $S/k_B = \partial (T \ln\Xi)/\partial T = \ln \Xi +
\Xi^{-1} T \partial \Xi/\partial T = \ln \Xi - \Xi^{-1} \beta \partial
\Xi/\partial \beta = \ln\Xi + \langle \beta(\Ham-\mu N)\rangle =
-\langle \ln f_0 \rangle$, see Refs.~\cite{eckert2020,
  eckert2023fluctuation} for related argumentation in the context of
spatially resolved measures of entropic and further equilibrium
fluctuations.

Using the functional form \eqref{EQOmegaFromEvaluationAtTheMinimizers}
together with the splitting
\eqref{EQmetaentropyOmegaViaEnergyEntropyAPX} we obtain
\begin{align}
  S &= -\frac{\partial \Omega[\rho,G]}{\partial T}\Big|_{\rho,G}\\
  &= S[\rho,G] \label{EQentroyResuls},
\end{align}
where the result certainly meets expectations {\ms and it is
  consistent with the standard definition
  \eqref{EQentropyFromOmega0}}. The simplicity of \eqr{EQentroyResuls}
arises from the vanishing of each of two additional terms that are
generated by the functional chain rule: $\int d\rv [\delta
  \Omega[\rho,G]/\delta \rho(\rv)]\partial \rho(\rv)/\partial T = 0$
and $\int dr [\delta \Omega[\rho,G]/\delta G(r)]/\partial
G(r)/\partial T=0$, as follows from each integrand vanishing due to
the respective minimization conditions \eqref{EQentropyEL1rawAPX} and
\eqref{EQentropyEL2rawAPX}. As an aside, the parametric temperature
derivative $\chi_T(\rv) = \partial \rho(\rv)/\partial T$ is the local
thermal susceptibility
\cite{eckert2020,eckert2023fluctuation,coe2022pre} that complements
the local compressibility $\chi_\mu(\rv) = \partial\rho(\rv)/\partial
\mu$ \cite{coe2022pre, stewart2012pre, evans2015jpcm, evans2019pnas,
  coe2022prl, coe2023, wilding2024, tarazona1985interface}; we return
to the relationship of these local measures of fluctuations with the
present framework in Sec.~\ref{SECentropyFluctuationProfiles} below.

To address the ideal gas, we revert to the standard classical density
functional setup of working with the grand potential density
functional $\Omega[\rho]$, where one can following similar lines of
argumentation as above. Exploiting that $\delta
\Omega[\rho]/\delta\rho(\rv)=0$ at the density functional minimum, one
can conclude that the nonvanishing contributions in the thermodynamic
relationship \eqref{EQentropyFromOmega0} are:
\begin{align}
  S = -\frac{\partial}{\partial T}
  (F_\rmid[\rho] + F_\rmexc[\rho, \beta \phi]),
  \label{EQentropySAsParametricDerivative}
\end{align}
where both $\rho(\rv)$ and $\phi(r)$ are kept fixed when performing
the parametric temperature derivative.  Here $F_\rmid[\rho]$ is the
standard intrinsic ideal gas Helmholtz free energy functional
(specified below) and $F_\rmexc[\rho,\beta\phi]$ is the corresponding
excess free energy functional; we have made the functional dependence
on the (scaled) pair potential $\beta\phi(r)$ explicit in the notation
\cite{kampa2024meta, kampa2026pairmatching, kampa2026spherical}. We
restrict ourselves to pairwise interacting systems and recall that
classical density functional theory applies to general many-body
interparticle interactions; see Ref.~\cite{bui2026abinitio} for
striking {\it ab initio} applications to molecular systems where the
Hamiltonian arises from treating the underlying electronic quantum
mechanics.

From the first term in the sum
\eqref{EQentropySAsParametricDerivative} one obtains the ideal
one-body entropy functional $S_{\rm id}[\rho]$ in the following
straightforward way. We start with the explicit form of the ideal free
energy density functional \cite{evans1979, hansen2013},
\begin{align}
  F_\rmid[\rho] 
  &= k_BT \int d\rv \rho(\rv)\big[\ln(\rho(\rv)\Lambda^d)-1\big],
  \label{EQentropyFidExplicit}
\end{align} 
where the thermal de Broglie wavelength \cite{hansen2013} is given by
$\Lambda = h/\sqrt{2\pi m k_BT}$.  From the product rule one can
conclude:
\begin{align}
  S_{\rm id}[\rho] &=
  -\frac{\partial F_\rmid[\rho]}{\partial T}\\
  &= -\frac{F_\rmid[\rho]}{T} 
  - k_BT \int d\rv \rho(\rv)
  \frac{d}{\Lambda}\frac{\partial\Lambda}{\partial T}\\
  &= -\frac{F_\rmid[\rho]}{T}
  + k_B \frac{d}{2} \int d\rv\rho(\rv)
  \label{EQentropySidOnePreliminary}\\
  &= -k_B \int d\rv \rho(\rv)
  \big[\ln\big(\rho(\rv)\Lambda^d\big) - (2+d)/2\big],
  \label{EQentropySidOneDefinition}
\end{align}
where the simplification \eqref{EQentropySidOnePreliminary} follows
from $\partial \Lambda / \partial T = -\Lambda/(2T)$; then inserting
the ideal free energy functional~\eqref{EQentropyFidExplicit} yields
upon simplifying the result \eqref{EQentropySidOneDefinition}.
Using the density functional form~\eqref{EQentropySidOneDefinition} in
the decomposition
\begin{align}
  F_\rmid[\rho] &= K[\rho] - TS_\rmid[\rho],
  \label{EQFidViaKandSid}
\end{align}
yields the classical mean kinetic energy density functional by
re-arranging, $K[\rho] = F_\rmid[\rho] + T S_{\rm id}[\rho]$, as
\begin{align}
  K[\rho] 
  &=  \frac{k_BTd}{2}\int d\rv \rho(\rv),
  \label{EQentropyKineticEnergyFunctional}
\end{align}
which is a result that again meets expectations.  We hence have
identified the separate entropic ideal
\eqref{EQentropySidOneDefinition} and kinetic
\eqref{EQentropyKineticEnergyFunctional} density functionals. These
are used when tracking the individual contributions to the grand
potential metadensity functional
\eqref{EQentropyOmegaDecompositionAPX} and in the definition of the
entropic splitting~\eqref{EQentropySidSexcSplittingAPX} into ideal and
excess contributions.

{\ms

It is instructive to give an alternative derivation of the ideal
functionals \eqref{EQentropySidOneDefinition}
and~\eqref{EQentropyKineticEnergyFunctional} via explicit calculation,
carried out on the many-body level as follows.  The interparticle
interaction potential vanishes for the ideal gas, $u(\rv^N)=0$, such
that the ideal version of the
Hamiltonian~\eqref{EQentropyHamiltonianAPX}, as denoted by
$H_{N,\rmid}$ in the following, consists merely of sums over decoupled
single-particle (kinetic and external) energy contributions.  Thus the
equilibrium measure factorizes,
$f_{0}(\rv^N,\pv^N)=(1/\Xi_\rmid)\Pi_{i=1}^N
(\e^{-\beta\pv_i^2/(2m)}\e^{-\beta( V_\rmext(\rv_i)-\mu)})$, with the
grand partition sum of the ideal gas $\Xi_\rmid = \Tr \e^{-\beta
  (H_{N,\rmid}-\mu N)} = \Tr f_0 \Xi_\rmid = \e^{Z_1}$ and the
single-particle partition sum $Z_1 =\int d\rv \e^{-\beta(
  V_\rmext(\rv)-\mu)}/\Lambda^d$.  We recall both the grand canonical
trace as $\Tr\,\cdot\, = \sum_{N=0}^\infty (h^{dN}N!)^{-1} \int d\rv^N
d\pv^N\,\cdot\,$ and the thermal wavelength as $\Lambda= h/\sqrt{2\pi
  m k_BT}$. Based on the
definition~\eqref{EQentropyMerminFunctionalDefinitionAPX} of the
Mermin functional, by explicit calculation one finds that
$\Omega_M[f_0] = K[\rho] - T S_\rmid[\rho] + \int d\rv
\rho(\rv)V_\rmext(\rv)$, with equilibrium density profile $\rho(\rv) =
\Tr f_0 \hat\rho(\rv) = \e^{-\beta (V_\rmext(\rv) - \mu)}/\Lambda^d$,
the classical ideal entropy functional $S_\rmid[\rho]$
\eqref{EQentropySidOneDefinition}, and the classical kinetic energy
functional $K[\rho]$ \eqref{EQentropyKineticEnergyFunctional}.
Explicit calculation also yields the equivalence of the respective
functional and many-body forms: $K[\rho] = \Tr f_0 \sum_i
\pv_i^2/(2m)$ and $S_\rmid[\rho]= -k_B \Tr f_0 \ln f_0$, which
confirms the physical interpretation of both density functionals.

}

{\ms

For completeness, we return to mutually interacting systems and
re-consider the constrained search definition
\eqref{EQentropyViaConstrainedMaximizationAPX} of the entropy
functional $S[\rho,G]$.  Performing the functional maximization with
respect to the dependence on the momentum degrees of freedom yields
the Maxwell distribution, such that $f(\rv^N,\pv^N) =
f_c(\rv^N)\e^{-\beta \sum_{i=1}^N\pv_i^2/(2m)}$.  Thus the phase space
momentum integrals are Gaussian and can be performed explicitly.  The
remaining nontrivial position coupling is captured by the
configurational trace, defined as $\Tr_c \,\cdot\, = \sum_{N=0}^\infty
(\Lambda^{dN} N!)^{-1} \int d\rv^N \,\cdot\,$.  
Thus one obtains the following configurational definition:
\begin{align}
  S[\rho,G] &= \max_{f_c\to \rho,G}\Big[
  -k_B \Tr_c f_c\ln f_c\Big],
  \label{EQentropyViaConstrainedMaximizationConfigurational}
\end{align}
where the two explicit kinetic energy contributions in the full phase
space form~\eqref{EQentropyViaConstrainedMaximizationAPX} have
cancelled.

We next provide an in-depth exploration of details and of consequences
that follow from the entropy functional point of view.

}

\section{Entropy functional framework}
\label{SECentropyFunctionalFramework}

The entropy functional is obtained from an entropic constrained search
method, as shown in Sec.~\ref{SECentropyAppendixOmegaDerivation}.
Partial functional minimization then recovers standard classical
density functional theory, as described in
Sec.~\ref{SECrelationshipWithStandardDFT}.  An intensive global pair
distribution function $\go(r)$ is used as a functional argument
alternative to $G(r)$, as is developed in
Sec.~\ref{SECchangeOfVariables}.  Entropic functional line integrals
are described in Sec.~\ref{SECentropyFunctionalIntegration}.

\subsection{Entropic constrained search}
\label{SECentropyAppendixOmegaDerivation}

The simple case of the ideal gas, described in
Sec.~\ref{SECidealFunctionals}, permits to obtain an exact solution in
terms of analytical density functionals for (classical) kinetic energy
and entropy. Arguably, one cannot hope to obtain an analytical
solution for the excess entropy functional. The present formal
developments are nevertheless important, as they establish existence,
uniqueness, and universality of this object. This has consequences for
subsequent developments of exact sum rules and of functional
interrelationships, for setting up functional machine learning
schemes, and for potentially making progress on the basis of
analytical approximations that go beyond the simplest second-virial
(Sec.~\ref{SECpairEntropyApproximation}) and mean-field
(Sec.~\ref{SECmeanField}) approaches. We recall that the arguably most
successful free energy density functional, the fundamental measure
theory for hard spheres \cite{rosenfeld1989}, consists of only
(thermally scaled) entropy, as the mean interparticle energy vanishes
for hard core systems.

We base the entropic metadensity functional construction on a
generalization of Levy's constrained search \cite{levy1979,
  dwandaru2011, schmidt2026entropyShort} and here provide details that
flesh out the overview given in Sec.~\ref{SECoverview} above.  The
relevant metadensity functional map is also proven via the alternative
standard Mermin-Evans {\it reductio ad absurdum} argument in
Sec.~\ref{SECentropyProofByContradiction}.

We insert the explicit form of Mermin's functional
\eqref{EQentropyMerminFunctionalDefinitionAPX} into the constrained
search \eqref{EQentropyOmegaFunctionalFromMerminAPX}, which gives:
\begin{align}
  \beta\Omega[\rho,G] &= \min_{f\to\rho,G}
  \Tr f (\ln f + \beta \Ham - \beta \mu N)
  \\&=
  \min_{f\to\rho,G} \Tr f
  \Big(\ln f +
   \beta \sumi \frac{\pv_i^2}{2m} 
  \notag\\&\qquad\qquad\qquad
  + \int_0^\infty dr \hat G(r) \beta \phi(r)
    \notag\\&\qquad\qquad\qquad
    + \int d\rv \hat\rho(\rv) \beta V_\rmext(\rv) 
    - \beta\mu N\Big),
  \label{EQentropyAppendixConstrainedSearch}
\end{align}
where in the second step we have written out the Hamiltonian $\Ham$ in
the re-written form \eqref{EQentropyHamiltonianRewrittenAPX}. One can
observe that the trace over phase space commutes both with the
position integral over $\rv$ and with the pair distance integral
over~$r$; we consider the corresponding terms separately in the
following.

First, the interparticle potential energy [third term in
  \eqr{EQentropyAppendixConstrainedSearch}] can be re-written as:
\begin{align}
  \Tr f \int_0^\infty
  dr\hat G(r) \beta \phi(r) 
  &= \int_0^\infty dr [\Tr f \hat G(r)] \beta\phi(r)  \\
  &=  \int_0^\infty dr  G(r) \beta\phi(r),
  \label{EQentropyAppendixInterparticleEnergy}
\end{align}
where we have first moved the trace and the trial distribution
function $f$ inside of the pair distance integral and then have used
the constraint \eqref{EQentropyConstraint2APX} to identify the trial
pair reduction function, $G(r)=\Tr f \hat G(r)$.

Secondly and similarly, the external potential energy [fourth term in
  \eqr{EQentropyAppendixConstrainedSearch}] is given in standard form
\cite{levy1979, dwandaru2011, evans1979, hansen2013} by:
\begin{align}
  \Tr f \int d\rv \hat \rho(\rv) \beta V_\rmext(\rv)
  &=
  \int d\rv [\Tr f \hat \rho(\rv)] \beta V_\rmext(\rv)  \\
  &= \int d\rv \rho(\rv) \beta V_\rmext(\rv),
  \label{EQentropyAppendixExternalEnergy}
\end{align}
where we have first interchanged the position integral and the trace
and then have used the constraint \eqref{EQentropyConstraint1APX} to
identify the trial density profile $\rho(\rv) = \Tr f \hat\rho(\rv)$.

The further chemical potential contribution [very last term in
  \eqr{EQentropyAppendixConstrainedSearch}] is:
\begin{align}
  -\Tr f \beta\mu N &= -\beta \mu  \Tr f \int d\rv \hat\rho(\rv)\\
  &= -\beta \mu \int d\rv \rho(\rv),
  \label{EQentropyAppendixChemicalEnergy}
\end{align}
where we have first expressed the particle number as $N=\int d\rv
\hat\rho(\rv)$, and then moved the trace and trial distribution inside
of the position integral to again identify the trial density profile
$\rho(\rv)$ via the constraint \eqref{EQentropyConstraint1APX}.

A crucial step in the constrained search construction is to recognize
that all three results \eqref{EQentropyAppendixInterparticleEnergy},
\eqref{EQentropyAppendixExternalEnergy}, and
\eqref{EQentropyAppendixChemicalEnergy} do {\it not} change when
changing $f(\rv^N, \pv^N)$ upon keeping the prescribed forms of
$\rho(\rv)$ and of $G(r)$ fixed, as is required to satisfy the joint
constraints \eqref{EQentropyConstraint1APX} and
\eqref{EQentropyConstraint2APX}. Hence the corresponding contributions
to the grand potential (interparticle potential, external potential,
and chemical potential) can be taken outside of the constrained
search~\eqref{EQentropyAppendixConstrainedSearch} and thus one
obtains:
\begin{align}
  \beta \Omega[\rho, G] &=
  \min_{f\to\rho,g} \Tr f \Big(\ln f
    +\beta\sumi\frac{\pv_i^2}{2m} \Big)
  \notag\\&\quad
  +\int_0^\infty dr G(r) \beta\phi(r)
  \notag\\&\quad
  +\int d\rv \rho(\rv) [\beta V_\rmext(\rv)-\beta \mu].
  \label{EQentropyOmegaSplittingAppendix}
\end{align}
The result \eqref{EQentropyOmegaSplittingAppendix} constitutes the
grand potential splitting \eqref{EQentropyOmegaDecompositionAPX} upon
identifying the first term on the right hand side via the definition
\eqref{EQentropyViaConstrainedMaximizationAPX} as the scaled sum
$\beta(-TS[\rho,G]+K[\rho])$.  The multiplication by $-k_B<0$ turns
the minimization \eqref{EQentropyOmegaDecompositionAPX} into the
maximization \eqref{EQentropyViaConstrainedMaximizationAPX}, which is
an arguably more natural convention in the present entropy context.
We describe an equivalent reduced free energy formulation, which
remains formulated via minimization, in
Sec.~\ref{SECreducedFreeEnergyPicture} below.

\subsection{Partial functional minimization}
\label{SECrelationshipWithStandardDFT}

Standard classical density functional theory, as recapped in the
Sec.~\ref{SECcdftOverview}, is based on the density profile
$\rho(\rv)$ as its (single) variational variable. This is generated
from the density constraint \eqref{EQentropyConstraint1APX} within the
constrained search method \cite{dwandaru2011}. The present entropic
framework is extended via the pair distance distribution constraint
\eqref{EQentropyConstraint2APX} and we demonstrate in the following
how the standard approach is recovered from the present entropic
framework.

It is useful to first flesh out the structure of standard classical
density functional theory, as summarized in
Sec.~\ref{SECcdftOverview}. Upon making the functional dependence on
the pair potential explicit, the standard Euler-Lagrange equation
\eqref{EQcdftEulerLagrange}, see Refs.~\cite{hansen2013, evans1979,
  evans1992, evans2016}, can be written as:
\begin{align}
  c_1(\rv;[\rho,\beta\phi]) = \ln(\rho(\rv)\Lambda^d)
  + \beta V_\rmext(\rv) - \beta\mu.
  \label{EQentropyEulerLagrangeStandardDFT}
\end{align}
Here $c_1(\rv;[\rho,\beta\phi])$ the standard one-body direct
correlation functional
\eqref{EQcdftOneBodyDirectCorrelationFunctional}, which is obtained
from the excess free energy density functional
$F_\rmexc[\rho,\beta\phi]$ via density functional differentiation
according to:
\begin{align}
  c_1(\rv;[\rho; \beta\phi]) &= 
  -\frac{\delta \beta  F_\rmexc[\rho,\beta\phi]}{\delta\rho(\rv)}
  \Big|_{\beta\phi}
  \label{EQc1Standard}
\end{align}
We have made the metadensity functional dependence on $\beta\phi(r)$
explicit in the notation and recall that this is accessible in
practice via functional machine learning~\cite{kampa2024meta,
  kampa2026pairmatching, kampa2026spherical}, such that functional
differentiation can be implemented on the basis of automatic
differentiation \cite{baydin2018autodiff, stierle2024autodiff}.

To make the connection with the present entropic framework, we first
re-write the grand potential metadensity
functional~\eqref{EQentropyOmegaDecompositionAPX} using the entropic
splitting into ideal and excess
contributions~\eqref{EQentropySidSexcSplittingAPX}. Then identifying
the ideal free energy density functional~\eqref{EQFidViaKandSid}
yields the following form of the grand potential metadensity
functional:
\begin{align}
  \Omega[\rho,G] &= 
   F_\rmid[\rho]  - TS_\rmexc[\rho,G] 
  +\int_0^\infty dr G(r) \phi(r)\notag\\
  &\quad
  +\int d\rv \rho(\rv)\big(V_\rmext(\rv)-\mu\big).
  \label{EQentropyOmegaDecompositionExcessAPX}
\end{align}

We next combine the scaled excess entropy contribution [second term in
  \eqr{EQentropyOmegaDecompositionExcessAPX}] with the interparticle
potential energy [third term in
  \eqr{EQentropyOmegaDecompositionExcessAPX}] to define an {\it
  extended} excess free energy metadensity functional,
\begin{align}
  {\cal F}_\rmexc([\rho, G,  \phi],T) &= - TS_\rmexc[\rho,G]
  +\int_0^\infty dr G(r) \phi(r),
  \label{EQentropyFexcMetaDensityFunctionalUnscaled}
\end{align}
Scaling the extended excess free energy functional
\eqref{EQentropyFexcMetaDensityFunctionalUnscaled} by thermal energy
allows one to combine the parametric dependence on temperature $T$
with the functional dependence on the pair potential $\phi(r)$. One
thus obtains the simpler form:
\begin{align}
  \beta {\cal F}_\rmexc[\rho, G, \beta \phi] &= - S_\rmexc[\rho,G]/k_B
  +\int_0^\infty dr G(r) \beta \phi(r),
  \label{EQentropyFexcMetaDensityFunctional}
\end{align}
In \eqr{EQentropyFexcMetaDensityFunctional} per construction the
functional dependence on the scaled pair potential $\beta \phi(r)$ is
linear and the nontrivial dependence on $\rho(\rv)$ and $G(r)$ is
encaspulated fully in $S_\rmexc[\rho,G]$.  
{\ms That no further parametric dependence on temperature $T$ occurs
  in the scaled excess free energy functional
  \eqref{EQentropyFexcMetaDensityFunctional} is consistent with the
  behaviour of the excess free energy metadensity functional
  \cite{kampa2024meta, kampa2026pairmatching, kampa2026spherical},
  where this has been validated numerically for a range of different
  short-ranged repulsive pair potentials upon varying the temperature;
  see Fig.~3 in Ref.~\cite{kampa2026spherical}.}

From the extended free energy
functional~\eqref{EQentropyFexcMetaDensityFunctional} one generates
the standard excess free energy density functional $\beta
F_\rmexc[\rho,\beta\phi]$ via partial minimization,
\begin{align}
  \beta F_\rmexc[\rho,\beta\phi]
  &= \min_G \beta {\cal F}_\rmexc[\rho,G,\beta\phi].
  \label{EQFexcFromPartialMinimizationAPX}
\end{align}
where both the density profile $\rho(\rv)$ and the scaled pair
potential $\beta\phi(r)$ are kept fixed when minimizing with respect
to $G(r)$.  The partial minimization form
\eqref{EQFexcFromPartialMinimizationAPX} follows from the joint
minimization principle \eqref{EQentropyMinimizationPrincipleAPX} upon
sequential ordering of the joint minimization according to
$\min_{\rho, G}\,\cdot\,=\min_\rho \min_G\,\cdot\,$. Then recognizing
that both the ideal free energy and the external potential
contribution are mere density functionals, which hence are independent
of $G(r)$ and do not affect the minimization over $G(r)$, leads to the
identity \eqref{EQFexcFromPartialMinimizationAPX}.  Then the grand
potential metadensity
functional~\eqref{EQentropyOmegaDecompositionExcessAPX} reduces to the
standard density functional form \eqref{EQcdftOmegaFunctional}:
\begin{align}
  \Omega[\rho] &= 
  F_\rmid[\rho] + F_\rmexc[\rho,\beta\phi]
  + \int d\rv \rho(\rv)\big( V_\rmext(\rv) -\mu \big),
  \label{EQmetaentropyOmegaFunctionalStandard}
\end{align}
where the sum of the first and second terms constitutes the total
intrinsic Helmholtz free energy density functional $F[\rho] =
F_\rmid[\rho] + F_\rmexc[\rho,\beta\phi]$.

The thus obtained standard free energy density functional splitting
\eqref{EQmetaentropyOmegaFunctionalStandard} should be contrasted with
the significantly higher functional resolution that the entropic
approach provides, see Eqs.~\eqref{EQmeanEnergyFunctionalAPX},
\eqref{EQmetaentropyOmegaViaEnergyEntropyAPX}, and
\eqref{EQentropySidSexcSplittingAPX}. The very moderate increase of
computational complexity over the standard density functional
dependence on~$\rho(\rv)$ consists of the mere additional functional
dependence on~$G(r)$. We re-iterate that $r$ remains a scalar
variable, irrespective of the spatial dimensionality $d$. 

The integral over all distances is over-extensive, $\int dr G(r) =
\langle N(N-1)/2\rangle$, as follows from the global reduction
operator \eqref{EQentropyInterparticlePotentialReWritingAPX}.  We
develop next an alternative and equivalent functional re-formulation
that circumvents this feature and can be more convenient in practice.

\subsection{Intensive functional variables}
\label{SECchangeOfVariables}

We start by addressing basic properties of the global pair reduction
function $G(r)$; the corresponding operator $\hat G(r)$ is given by
the definition \eqref{EQentropyGhatAPX}. For the ideal gas with
density profile $\rho(\rv)$ one has $G(r)=G_\rmid(r;[\rho])$ with the
explicit density functional:
\begin{align}
  G_\rmid(r;[\rho]) &= 
  \frac{1}{2}\int d\rv d\rv'\rho(\rv)\rho(\rv')
  \delta(r-|\rv-\rv'|),
  \label{EQentropyGiddy}
\end{align}
which follows from the two-body density factorizing trivially (here)
into the product of one-body densities.

In practice, when addressing non-ideal systems, it can be advantageous
to use the ideal form \eqref{EQentropyGiddy} as a reference and
instead of working with $G(r)$ to rather consider its deviation from
the ideal gas form \eqref{EQentropyGiddy}. We hence define the {\it
  global} pair distribution function $\go(r)$ as the following ratio:
\begin{align}
 \go(r)=\frac{G(r)}{G_\rmid(r;[\rho])},
  \label{EQentropySmallgAsRatio}
\end{align}
where $G_\rmid(r;[\rho])$ is the density functional
\eqref{EQentropyGiddy}.  {\ms We use the notation $\go(r)$ to
  distinguish the global version \eqref{EQentropySmallgAsRatio} from
  the standard homogeneous bulk fluid pair distribution function
  $g(r)$. Working with intensive variables can have the usual benefit
  of taking out the primary dependence on system size.}  Any
functional dependence on $G(r)$ then implies dependence on $\go(r)
G_\rmid(r;[\rho])$, such that the excess entropy functional
$S_\rmexc[\rho, G]=S_\rmexc[\rho,\go G_\rmid]$ transfers its
functional dependence from $\rho(\rv)$ and $G(r)$ to $\rho(\rv)$ and
$\go(r)$.

We investigate the changes in the formal theoretical structure upon
change of functional variables $\rho(\rv)$ and $G(r)$ to $\rho(\rv)$
and $\go(r)$. As the variable
transformation~\eqref{EQentropySmallgAsRatio} from $G(r)$ to $\go(r)$
depends on the density profile via the density functional
\eqref{EQentropyGiddy}, one has to monitor the corresponding induced
changes when building functional derivatives, as we lay out in the
following. The argumentation leads to the alternative form of the
Euler-Lagrange equations \eqref{EQentropyEL2alternative1} and
\eqref{EQentropyEL2alternative2} given below.

The following density functional derivative defines an alternative
first-order direct correlation functional:
\begin{align}
  c_{\rho|\go}(\rv;[\rho,\go]) &=
  \frac{\delta S_\rmexc[\rho,\go G_\rmid]/k_B}{\delta \rho(\rv)} \Big|_\go
  \label{EQentropyDirectcrhog}
\end{align}
where on the left hand side the subscript $\rho|\go$ indicates that
the functional derivative with respect to $\rho(\rv)$ is taken upon
fixing $\go(r)$ rather than $G(r)$; we recall that~$G(r)$ is kept
fixed in the density functional
derivative~\eqref{EQentropyDirectEntropic1APX} that defines the
entropic direct correlation functional $c_\rho(\rv;[\rho,G])$.

From the definition \eqref{EQentropyDirectcrhog} one obtains:
\begin{align}
  c_{\rho|\go}(\rv;[\rho,\go]) 
  &=c_\rho(\rv;[\rho,\go G_\rmid]) 
  \notag\\&\quad
  +\int_0^\infty dr' c_G(r';[\rho,\go G_\rmid])
  \notag\\&\qquad\qquad\qquad \times
  \go(r') \frac{\delta G_\rmid(r';[\rho])}{\delta \rho(\rv)},
  \label{EQentropyIntermediate1}
\end{align}
where in the re-writing \eqref{EQentropyIntermediate1} the first term
arises as the direct functional derivative with respect to $\rho(\rv)$
according to the definition \eqref{EQentropyDirectEntropic1APX}, i.e.,
upon keeping the second functional argument $G(r)$ fixed. The second
term in \eqr{EQentropyIntermediate1} is due to the functional chain
rule and we recall the entropic direct correlation functional
$c_G(r';[\rho,G])$ being given by the definition
\eqref{EQentropyDirectEntropic2APX} and that $G_\rmid(r;[\rho])$ is
the ideal gas global pair reduction functional \eqref{EQentropyGiddy}.

Based on the definition \eqref{EQentropyGiddy} of the ideal gas form
$G_\rmid(r;[\rho])$ one obtains straightforwardly its density
functional derivative as
\begin{align}
  \frac{\delta G_\rmid(r;[\rho])}{\delta\rho(\rv)} &=
  \int d\rv'\rho(\rv')\delta(r-|\rv-\rv'|).
\end{align}
This result allows one to express the second term in
\eqr{EQentropyIntermediate1}, upon carrying out the integral over
$r'$, as the following convolution integral:
\begin{align}
  (c_G \go \star \rho)(\rv)  &=
  \int d\rv' c_G(|\rv-\rv'|) \go(|\rv-\rv'|) \rho(\rv').
  \label{EQconvolution1}
\end{align}
Similarly, we define the convolution of the
product~$\beta\phi(r)\go(r)$, where we emphasize that the
multiplication of $\beta\phi(r)$ and $\go(r)$ is performed in real
space, with the density profile $\rho(\rv)$ as
\begin{align}
  (\beta\phi \go \star \rho)(\rv)  &=
  \beta \int d\rv' \beta\phi(|\rv-\rv'|) \go(|\rv-\rv'|) \rho(\rv').
  \label{EQconvolution2}
\end{align}
From the `distance' Euler-Lagrange equation
\eqref{EQentropyEL2rhoGAPX}, i.e., $c_G(r;[\rho,G]) = \beta\phi(r)$,
one can conclude the equality of the two convolution integrals
\eqref{EQconvolution1} and \eqref{EQconvolution2}:
\begin{align}
  (c_G \go \star \rho)(\rv)  &=
  (\beta\phi \go \star \rho)(\rv).
  \label{EQentropyIntermediate3}
\end{align}
Using the convolution notation \eqref{EQconvolution1} allows one to
express the direct correlation functional identity
\eqref{EQentropyIntermediate1} in the following succinct form:
\begin{align}
  c_{\rho|\go}(\rv;[\rho,\go])  &=
  c_\rho(\rv;[\rho,G]) +  (c_G \go \star \rho)(\rv),
  \label{EQentropyIntermediate2}
\end{align}
which can be re-written upon using \eqr{EQentropyIntermediate3} as
\begin{align}
  c_{\rho|\go}(\rv;[\rho,\go])  &=
  c_\rho(\rv;[\rho,G])  +
  (\beta\phi \go \star \rho)(\rv).
  \label{EQentropyEntropicDirectCorrelationRelationAgain}
\end{align}

To complement the direct correlation functional
\eqref{EQentropyDirectcrhog}, we also define
\begin{align}
  c_g(r;[\rho,\go])
  &=
  \frac{\delta S_\rmexc[\rho,\go G_\rmid]/k_B}{\delta \go(r)}\Big|_\rho
  \label{EQdirectconeAlternative2}  \\
  &= G_\rmid(r;[\rho]) c_G(r;[\rho,\go G_\rmid]),
  \label{EQdirectconeAlternative3}
\end{align}
where the latter form follows from the chain rule and the fact that
$G_\rmid(r;[ \rho])$ is independent of $\go(r)$; see the definition
\eqref{EQentropyGiddy} of the ideal gas pair reduction
functional. Using the results
\eqref{EQentropyEntropicDirectCorrelationRelationAgain} and
\eqref{EQdirectconeAlternative3} allows one to re-write the
Euler-Lagrange equations~\eqref{EQentropyEL1rhoGAPX} and
~\eqref{EQentropyEL2rhoGAPX} as
\begin{align}
  c_{\rho|\go}(\rv;[\rho,\go]) &=  
  \ln\rho(\rv) 
  + (\beta\phi \go\star\rho)(\rv)
  \notag\\&\quad
  + \beta V_\rmext(\rv) - \beta\mu,
  \label{EQentropyEL2alternative1}\\
  c_g(r;[\rho,\go]) &= G_\rmid(r;[\rho]) \beta \phi(r)
  \label{EQentropyEL2alternative2},
\end{align}
which constitutes the desired formulation that is suitable for working
with $\rho(\rv)$ and $\go(r)$.

\subsection{Entropic functional integrals}
\label{SECentropyFunctionalIntegration}

We next apply the concept of functional line integration
\cite{evans1992, sammueller2023whyNeural} to the present entropy
functional derivatives. One obtains the excess entropy according to
one of the following functional integrals; these are inverses of the
respective functional derivatives. Starting with the density
functional derivative \eqref{EQentropyDirectEntropic1APX} one has the
following functional line integral:
\begin{align}
  S_\rmexc[\rho,G] &= 
  {\ms S_\rmexc[0,G]} + 
  k_B \int d\rv \rho(\rv) \int_0^1 da c_\rho(\rv;[a\rho,G]),
  \label{EQentropySexcViaLineIntegral1}
\end{align}
where the density functional argument has the standard scaled form
$a\rho(\rv)$ with parameter $0\leq a \leq 1$, as is prescribed by the
integral boundaries. {\ms We expect} the lower boundary term to
vanish, $S_\rmexc[0,G]=0$ at vanishing density $\rho(\rv)\to 0$.
{\ms This expectation can be tested, in principle, by calculating the
  boundary term $S_\rmexc[0,G]$ via the general procedure described in
  the following, upon specializing this to the case $\rho(\rv)\to
  0$. Note that the corresponding boundary term $S_\rmexc[0,\go]$ will
  vanish strictly, when working with the alternative functional
  integral \eqref{EQdirectconeAlternative2} that is based on the
  change to the intensive functional variables of
  Sec.~\ref{SECchangeOfVariables}.}

Furthermore corresponding to the functional
derivative~\eqref{EQentropyDirectEntropic2APX} we have:
\begin{align}
  S_\rmexc[\rho,G] &= k_B
  \int_0^\infty dr \Delta G(r) \int_0^1 da c_G(r;[\rho,G_a]),
  \label{EQentropySexcViaLineIntegral2}
\end{align}
where $G_a(r) = (1-a)G_\rmid(r;[\rho]) + aG(r)$ interpolates between
the behaviour of the inhomogeneous ideal gas and that of the real
system. We have defined $\Delta G(r) = \partial G_a(r)/\partial a =
G(r) - G_\rmid(r;[\rho])$ and we recall $G_\rmid(r;[\rho])$ being
given by the definition \eqref{EQentropyGiddy}. The excess entropy
vanishes at the lower limit of the functional line integral,
$S_\rmexc[\rho,G_\rmid]=0$, as follows for the ideal gas.

When working alternatively with the pair of variational
fields~$\rho(\rv)$ and~$\go(r)$ then the following functional line
integrals correspond to \eqr{EQentropyDirectcrhog}:
\begin{align}
  S_\rmexc[\rho,G] &= k_B
  \int d\rv \rho(\rv) \int_0^1 da c_{\rho|\go}(\rv;[a \rho,\go]),
\end{align}
and to \eqr{EQdirectconeAlternative2}:
\begin{align}
  S_\rmexc[\rho,\go] &= k_B
  \int_0^\infty dr \big(\go(r)-1\big) \int_0^1 da c_\go(\rv;[\rho, \go_a]),
\end{align}
where $\go_a(r) = 1-a + a\go(r)$ is a parameterized global pair
distribution function. At the lower limit $\go_0(r)=1$ and at the
upper limit $\go_1(r)=\go(r)$, and the parametric derivative is
$\partial \go_a(r)/\partial a=\go(r)-1$.

\section{Entropic response and correlation}
\label{SECresponseAndCorrelation}

We address structural properties via entropic response and fluctuation
correlation functions that follow from (second-order) functional
differentiation in Sec.~\ref{SECsecondOrderResponseFunction}. {\ms
  Entropic second-order direct correlation functionals are introduced
  in Sec.~\ref{SECsecondOrderDirect}.}  Coupled exact entropic
Ornstein-Zernike equations interrelate the probabilistic correlation
functions with the entropic direct correlation functions, as is
presented in Section~\ref{SECOrnsteinZernikeExcessEntropy}.  The
formal connection with local (one-body) fluctuation profiles is
described in Sec.~\ref{SECentropyFluctuationProfiles}.

\subsection{Second-order response functions}
\label{SECsecondOrderResponseFunction}

The correlation function of density fluctuations $H_2(\rv,\rv')$ is
one of the most fundamental two-body functions that characterizes
spatial structure \cite{evans1979, hansen2013, schmidt2022rmp}. When
expressed via averages, the definition is $H_2(\rv,\rv') =
\cov(\hat\rho(\rv), \hat\rho(\rv'))$, where $\hat\rho(\rv)$ is the
density operator \eqref{EQdensityOperatorAPX} and the covariance is
defined below.  Together with knowledge of the density profile,
$H_2(\rv,\rv')$ determines the two-body density distribution
$\rho_2(\rv,\rv')$ and, for bulk fluids, the pair distribution
function and the structure factor. Furthermore the response of the
local density $\rho(\rv)$ against changes in the (thermally scaled)
external potential $\beta V_\rmext(\rv')$ is described via the exact
functional derivative relationship $H_2(\rv,\rv')=-\delta
\rho(\rv)/\delta \beta V_\rmext(\rv')$.

The covariance of two phase space functions $\hat A$ and $\hat B$ is
defined in the standard way as $\cov(\hat A, \hat B) = \langle \hat A
\hat B \rangle - \langle \hat A \rangle \langle \hat B \rangle$, where
we recall the angles to denote the thermal equilibrium average,
$\langle\,\cdot\,\rangle = \Tr f_0\,\cdot\,$.
As the entropic formulation treats both $\hat\rho(\rv)$ and $\hat
G(r)$ on an equal footing it is natural to define corresponding
covariances of pairs of these operators.  We change the notation
$H_2(\rv,\rv') = H_{\rho\rho}(\rv,\rv')$ and introduce similarly
$H_{\rho G}(\rv,r')$, $H_{G\rho}(r,\rv')$, and $H_{GG}(r,r')$,
together with associated total correlation functions
$h_{\rho\rho}(\rv,\rv')$, $h_{\rho G}(\rv,r')$, $h_{G\rho}(r,\rv')$,
and $h_{GG}(r,r)$. The specific definitions are as follows:
\begin{align}
  H_{\rho\rho}(\rv,\rv') &=
  -\frac{\delta \rho(\rv)}{\delta \beta V_\rmext(\rv')}\Big|_{\phi}
  =  \cov(\hat\rho(\rv),\hat\rho(\rv'))\notag\\
  &= \rho(\rv) \delta(\rv-\rv') + \rho(\rv)\rho(\rv')h_{\rho\rho}(\rv,\rv'),
  \label{EQentropyHrhorhoDefinition}
  \\
  H_{\rho G}(\rv,r') &=  
  -\frac{\delta\rho(\rv)}{\delta \beta\phi(r')}\Big|_{V_\rmext}
  = \cov(\hat\rho(\rv), \hat G(r'))\notag\\
  &= \rho(\rv) G(r') h_{\rho G}(\rv,r')
  = H_{G\rho}(r',\rv),
  \label{EQentropyHrhoGdefinition}
\end{align}
{\ms 
where the thermodynamic parameters $\mu$ and $T$ are kept fixed upon
building the functional derivatives.

The functional derivative form \eqref{EQentropyHrhorhoDefinition}
implies the standard interpretation \cite{hansen2013, evans1979} of
$H_2(\rv,\rv') = H_{\rho\rho}(\rv,\rv')$ measuring the response of the
local density at position~$\rv$ against a localized infinitesimal
decrease (cf.~the minus sign) of the scaled external potential at
position $\rv'$. This relationship is reciprocal such that the two
positions can be interchanged, as is apparent from the (trivial)
exchange symmetry of the covariance
\eqref{EQentropyHrhorhoDefinition}.

Similarly, the functional derivative~\eqref{EQentropyHrhoGdefinition}
implies the analogous interpretation of $H_{\rho G}(\rv,r')$ as
measuring the response of the local density at position $\rv$ against
a decrease (cf.~again the minus sign) in the value of the pair
potential at interparticle distance $r'$. Per construction this change
acts across the entirety of the system with fixed external potential
$V_\rmext(\rv)$ and fixed thermodynamic parameters $\mu$ and $T$,
which is inline with the physical interpretation of response. That the
resulting response function $H_{\rho G}(\rv,r')$ is identical to the
covariance \eqref{EQentropyHrhoGdefinition} of the local density
$\hat\rho(\rv)$ and the global pair reduction operator $\hat G(r')$ is
a nontrivial fact, which manifests the general mechanism of creating
meaningful averages via differentiation, whether performed
parametrically or in this case as functional derivatives.

}

The entirety of second-order correlation functions comprises a formal
$2\times 2$ matrix with the following two further elements:
\begin{align}
  H_{G\rho}(r,\rv') &= 
  -\frac{\delta G(r)}{\delta \beta V_\rmext(\rv')}\Big|_\phi
  =  \cov(\hat G(r),\hat\rho(\rv'))
  \notag\\ &
  = G(r)\rho(\rv') h_{G\rho}(r,\rv')
  = H_{\rho G}(\rv',r),
  \label{EQentropyHGrhoDefinition}
  \\
  H_{GG}(r,r') &= 
  - \frac{\delta G(r)}{\delta \beta\phi(r')}\Big|_{V_\rmext}
  = \cov(\hat G(r), \hat G(r'))
  \notag\\ &
  = G(r)\delta(r-r') + G(r)G(r') h_{GG}(r,r'),
  \label{EQentropyHGGdefinition}
\end{align}{\ms 
where again $\mu,T$ are fixed upon differentiating.

Each of the two functional derivatives
\eqref{EQentropyHGrhoDefinition} and \eqref{EQentropyHGGdefinition}
carries again physical interpretation as response. Here
$H_{G\rho}(r,\rv')$ implies the change of the global pair distance
histogram $G(r)$ at interparticle distance $r$ in response to a local
decrease of the scaled external potential $\beta V_\rmext(\rv')$ at
position $\rv'$. Similarly, $H_{GG}(r,r')$ measures the change of
$G(r)$ in response to the decrease of the pair potential at
interparticle distance~$r$, thus autocorrelating the global pair
reduction operator $\hat G(r)$ with itself.

}

The second equality in each of the
identities~\eqref{EQentropyHrhorhoDefinition}--\eqref{EQentropyHGGdefinition}
relates the respective functional derivative with the associated
covariance, as can be shown from explicit calculation using standard
argumentation \cite{hansen2013, evans1979, schmidt2022rmp,
  eckert2023fluctuation}.  Similar to the situation in the standard
theory \cite{evans1979, hansen2013, schmidt2022rmp} the response
functions feature in associated Ornstein-Zernike equations, to which
we turn after first laying out the direct correlation structure on the
level of two functional derivatives.

{\ms

\subsection{Second-order direct correlation functionals}
\label{SECsecondOrderDirect}

The standard integral equation theories for homogeneous liquid state
properties relate the pair correlation function to the bulk two-body
direct correlation function $c_2^b(r)$ \cite{hansen2013}; see
Ref.~\cite{pihlajamaa2024closures} for recent work. The density
functional perspective provides a generative mechanism for $c_2^b(r)$
via the second-order functional derivative of the (scaled) excess free
energy functional, $c_2^b(|\rv-\rv'|) = -\delta \beta
F_\rmexc[\rho]/[\delta \rho(\rv)\delta\rho(\rv')]|_{\rho=\rho_b}$,
where the notation indicates the evaluation at the bulk density
$\rho_b=\rm const$.
Equivalently one can start from the general one-body direct
correlation functional $c_1(\rv;[\rho])$, which we recall features
prominently in the standard Euler-Lagrange equation
\eqref{EQcdftEulerLagrange} and is given as the first-order density
functional derivative~\eqref{EQcdftOneBodyDirectCorrelationFunctional}
of $-\beta F_\rmexc[\rho]$.  Further functional differentiation then
yields the two-body direct correlation functional
$c_2(\rv,\rv';[\rho]) = \delta c_1(\rv;[\rho])/\delta\rho(\rv')$,
where we have kept the density functional dependence general. The
dependence on temperature is suppressed in the notation, as is the
metadensity functional dependence on the pair (and possibly
higher-body) interparticle interaction potential(s).

The present entropic approach suggests a corresponding derivative
structure to obtain entropic second-order direct correlation
functionals.  These will turn out to play a similar, yet extended,
role over the standard two-body direct correlation functional
$c_2(\rv,\rv';[\rho])$, as will become apparent later both in the
entropic Ornstein-Zernike approach
(Sec.~\ref{SECOrnsteinZernikeExcessEntropy}) and in the
hyperfunctional theory for general observables
(Sec.~\ref{SEChyperfunctionalEntropy}).

We start by recalling the entropic first-order direct correlation
functionals $c_\rho(\rv;[\rho,G])$ and $c_G(r;[\rho,G])$ as being
given, respectively, by the first-order functional
derivatives~\eqref{EQentropyDirectEntropic1APX}
and~\eqref{EQentropyDirectEntropic2APX}. Differentiating each of these
functionals with respect to one of its functional arguments yields the
following definitions:
\begin{align}
  c_{\rho\rho}(\rv,\rv';[\rho,G]) &= 
  \frac{\delta c_\rho(\rv;[\rho,G])}{\delta \rho(\rv')}
  \Big|_G,
  \label{EQentropySecondOrderDirect1}\\
  c_{\rho G}(\rv,r';[\rho,G]) &=
  \frac{\delta c_\rho(\rv;[\rho,G])}{\delta G(r')}
  \Big|_\rho,
  \label{EQentropySecondOrderDirect2}\\
  c_{G\rho}(r,\rv';[\rho,G]) &=
  \frac{\delta c_G(r;[\rho,G])}{\delta \rho(\rv')}
  \Big|_G,
  \label{EQentropySecondOrderDirect3}\\
  c_{GG}(r,r';[\rho,G]) &=
  \frac{\delta c_G(r;[\rho,G])}{\delta G(r')}
  \Big|_\rho.
  \label{EQentropySecondOrderDirect4}
\end{align}
As before, the subscript of the vertical bar indicates the functional
argument that is kept constant under the derivative. The two
second-order `cross' definitions \eqref{EQentropySecondOrderDirect2}
and \eqref{EQentropySecondOrderDirect3} possess an argument exchange
symmetry, $c_{\rho G}(\rv,r';[\rho,G]) = c_{G\rho}(r',\rv;[\rho,G])$,
as follows from interchanging the order of the implied second
functional derivatives of $S_\rmexc[\rho,G]$.  }

\subsection{Entropic Ornstein-Zernike equations}
\label{SECOrnsteinZernikeExcessEntropy}

We present Ornstein-Zernike equations that arise from the entropy
functional structure. The derivation is based on building first-order
functional derivatives of the Euler-Lagrange equations
\eqref{EQentropyEL1rhoGAPX} and \eqref{EQentropyEL2rhoGAPX}. {\ms
This method is general and it provides a systematic means to obtain
exact inhomogeneous second-order Ornstein-Zernike relations. Examples
include the derivations of the standard equilibrium inhomogeneous
two-body relationship~\cite{schmidt2022rmp}, the nonequilibrium
Ornstein-Zernike equations \cite{brader2013noz, brader2014noz,
  schmidt2022rmp}, the (one-body) fluctuation Ornstein-Zernike
relations \cite{eckert2020, eckert2023fluctuation}, the
hyper-Ornstein-Zernike relationships for the fluctuation structure of
general observables \cite{sammueller2024hyperDFT,
  sammueller2024whyhyperDFT, sammueller2024multihyperDFT} as is
re-examined below in Sec.~\ref{SEChyperfunctionalEntropy}, as well as
the meta-Ornstein-Zernike equation \cite{kampa2024meta,
  kampa2026pairmatching, kampa2026spherical}.

}

In the present extended entropy framework, each of the two
Euler-Lagrange equations \eqref{EQentropyEL1rhoGAPX} and
\eqref{EQentropyEL2rhoGAPX} is amenable to functional differentiation
with respect to either $\beta V_\rmexc(\rv')$ or $\beta\phi(r')$. Here
the primes indicate new variables, which are in general different from
the position $\rv$ and distance $r$ of the starting equations. Using
the functional chain rule, one obtains the following set of four
coupled exact entropic Ornstein-Zernike equations. {\ms The first pair
  of equations is obtained from differentiating the `position'
  Euler-Lagrange equation \eqref{EQentropyEL1rhoGAPX} with respect to
  either $\beta V_\rmext(\rv)$ or with respect to $\beta
  \phi(r)$. This yields, respectively: }
\begin{align}
  &  \int d\rv'' c_{\rho\rho}(\rv,\rv'') H_{\rho\rho}(\rv'',\rv')
  + \int dr'' c_{\rho G}(\rv,r'') H_{G\rho}(r'',\rv')
  \notag\\&\quad
  = \rho(\rv)^{-1} H_{\rho\rho}(\rv,\rv') - \delta(\rv-\rv'),
  \label{EQentropyOZ1}
  \\
  &  \int d\rv'' c_{\rho\rho}(\rv,\rv'') H_{\rho G}(\rv'',r')
  + \int dr'' c_{\rho G}(\rv,r'') H_{GG}(r'',r')
  \notag\\&\quad
  = \rho(\rv)^{-1} H_{\rho G}(\rv,r'),
\end{align}
{\ms with the second-order fluctuation correlation
  functions~\eqref{EQentropyHrhorhoDefinition}--\eqref{EQentropyHGGdefinition},
  and the second-order direct correlation
  functionals~\eqref{EQentropySecondOrderDirect1}--\eqref{EQentropySecondOrderDirect4};
  we have suppressed the functional dependence of the latter on
  $\rho(\rv)$ and $G(r)$ for notational clarity.}

{\ms The second pair of equations follows from treating the `distance'
  Euler-Lagrange equation \eqref{EQentropyEL2rhoGAPX} similarly, i.e.,
  by differentiating both sides of each equation with respect to
  either $\beta V_\rmext(\rv)$ or with respect to $\beta
  \phi(r)$. This strategy results in the following respective exact
  identities: }
\begin{align}
  &  \int d\rv'' c_{G\rho}(r,\rv'') H_{\rho\rho}(\rv'',\rv')
  + \int dr'' c_{GG}(r,r'') H_{G\rho}(r'',\rv')
  \notag\\&\quad
  = 0,
  \\
  &  \int d\rv'' c_{G\rho}(r,\rv'') H_{\rho G}(\rv'',r')
  + \int dr'' c_{GG}(r,r'') H_{GG}(r'',r')
  \notag\\&\quad
  = -\delta(r-r').
  \label{EQentropyOZ4}
\end{align}
Using the four total entropic correlation functions
$h_{\rho\rho}(\rv,\rv')$, $h_{\rho G}(\rv,r)$, $h_{G\rho}(r,\rv')$,
and $ h_{GG}(r,r')$, as introduced via
Eqs.~\eqref{EQentropyHrhorhoDefinition}--\eqref{EQentropyHGGdefinition},
the set of coupled
equations~\eqref{EQentropyOZ1}--\eqref{EQentropyOZ4} can be re-written
equivalently as:
\begin{align}
  &  c_{\rho\rho}(\rv,\rv') 
  + \int d\rv'' c_{\rho\rho}(\rv,\rv'') \rho(\rv'')h_{\rho\rho}(\rv'',\rv')
  \notag\\&\quad
  + \int dr'' c_{\rho G}(\rv,r'') G(r'')h_{G\rho}(r'',\rv')
  = h_{\rho\rho}(\rv,\rv'),
  \label{EQentropyOZ1alt}  \\
  & c_{\rho G}(\rv,r')
  + \int d\rv'' c_{\rho\rho}(\rv,\rv'') \rho(\rv'') h_{\rho G}(\rv'',r')
  \notag\\&\quad
  + \int dr'' c_{\rho G}(\rv,r'') G(r'') h_{GG}(r'',r')
  = h_{\rho G}(\rv,r'),
\end{align}
and furthermore:
\begin{align}
  & c_{G\rho}(r,\rv') 
  + \int d\rv'' c_{G\rho}(r,\rv'') \rho(\rv'') h_{\rho\rho}(\rv'',\rv')
  \notag\\&\quad
  + \int dr'' c_{GG}(r,r'') G(r'') h_{G\rho}(r'',\rv')
  = 0,
  \\
  & c_{GG}(r,r') 
  + \int d\rv'' c_{G\rho}(r,\rv'') \rho(\rv'') h_{\rho G}(\rv'',r')
  \notag\\&\quad
  + \int dr'' c_{GG}(r,r'') G(r'') h_{GG}(r'',r')
  = -\frac{\delta(r-r')}{G(r)}.
  \label{EQentropyOZ4alt}
\end{align}

{\ms 

The considerably large `depth of field' that the set of coupled
entropic Ornstein-Zernike equations
\eqref{EQentropyOZ1alt}--\eqref{EQentropyOZ4alt} provides should be
contrasted against the standard free-energy based version, which we
give in full form below, to illustrate the striking contrast in
complexity. The standard approach delivers merely the single
relationship \cite{hansen2013}:
\begin{align}
  c_2(\rv,\rv';[\rho]) +
  \int d\rv'' c_2(\rv,\rv'';[\rho])\rho(\rv'') h(\rv'',\rv')
  &= h(\rv,\rv'),
  \label{EQOrnsteinZernikeStandard}
\end{align}
where $h(\rv,\rv')$ is the inhomogeneous total pair correlation
function. We refer the reader to Ref.~\cite{schmidt2022rmp} for the
straightforward derivation of \eqr{EQOrnsteinZernikeStandard} based on
the present functional differentiation method applied in the free
energy-based context.}

\subsection{Localized fluctuation profiles}
\label{SECentropyFluctuationProfiles}

It is interesting to relate the present treatment of fluctuations with
the recent one-body fluctuation profiles \cite{coe2022pre,
  stewart2012pre, evans2015jpcm, evans2019pnas, coe2022prl, coe2023,
  wilding2024, tarazona1985interface, eckert2020,
  eckert2023fluctuation, coe2022pre}.
The one-body thermal susceptibility $\chi_T(\rv)$ introduced by Eckert
{\it et~al.}~\cite{eckert2020, eckert2023fluctuation} is defined as a
parametric derivative of the equilibrium one-body density profile with
respect to temperature, $\chi_T(\rv) = \partial \rho(\rv)/\partial T$.
One may express $\chi_T(\rv)$ alternatively via the phase space
covariance $\chi_T(\rv) = \cov(\hat\rho(\rv), \beta \Ham-\beta \mu
N)/T$. Due to the additive form of the Hamiltonian
\eqref{EQentropyHamiltonianRewrittenAPX}, the interparticle
contribution $\chi_{T,\rmint}(\rv)$ follows as
\cite{eckert2023fluctuation}:
\begin{align}
  \chi_{T,\rmint}(\rv) &= \cov(\hat\rho(\rv),\beta u(\rv^N))/T.
\end{align}
We recall $\hat\rho(\rv)$ as the density operator
\eqref{EQdensityOperatorAPX} and the interparticle interaction
potential $u(\rv)$ to be the additive contribution
\eqref{EQentropyInterparticlePotentialPairwiseAPX} to the Hamiltonian
\eqref{EQentropyHamiltonianRewrittenAPX}. Then re-writing $u(\rv^N)$
in the pairwise form
\eqref{EQentropyInterparticlePotentialReWritingAPX} introduces the
pair reduction function~$\hat G(r)$. We hence obtain:
\begin{align}
  \chi_{T,\rmint}(\rv) &= 
  \int_0^\infty dr \beta\phi(r) \cov(\hat\rho(\rv),\hat G(r))/T
  \\
  &= \int_0^\infty dr \beta\phi(r) H_{\rho G}(\rv,r)/T,
  \label{EQentropyChiTintRewritten}
\end{align}
where in the re-writing \eqref{EQentropyChiTintRewritten} we have
identified $H_{\rho G}(\rv,r) = \cov(\hat\rho(\rv),\hat G(r))$ as the
fluctuation correlation function~\eqref{EQentropyHrhoGdefinition} of
the local density $\hat\rho(\rv)$ and the global pair reduction
function $\hat G(r)$.

For completeness, we recall the relationship with the local
compressibility $\chi_\mu(\rv)$ \cite{coe2022pre, stewart2012pre,
  evans2015jpcm, evans2019pnas, coe2022prl, coe2023, wilding2024,
  tarazona1985interface}, which is given by $\chi_\mu(\rv) = \partial
\rho(\rv)/\partial \mu = \beta \cov(\hat\rho(\rv),N) = \beta \int
d\rv' H_{\rho\rho}(\rv,\rv')$, where we recall our notation
$H_{\rho\rho}(\rv,\rv') = H_2(\rv,\rv')$ for the standard two-body
correlation function of density
fluctuations~\eqref{EQentropyHrhorhoDefinition}.

{\ms

The fluctuations of a general physical quantity, which is given by a
`hyperobservable' $\hat A$ as a targeted phase space function, are
characterized by the position-resolved hyperfluctuation profile
$\chi_\rho^A(\rv)$ \cite{sammueller2024hyperDFT,
  sammueller2024whyhyperDFT, sammueller2024multihyperDFT}. We here
supplement its definition \eqref{EQchiAdensity} by the analogous pair
distance-resolved version~\eqref{EQchiAGhat}:
\begin{align}
  \chi_\rho^A(\rv) = \cov\big(\hat A, \hat\rho(\rv)\big),
  \label{EQchiAdensity}  \\
  \chi_G^A(r) = \cov\big(\hat A, \hat G(r)\big).
  \label{EQchiAGhat}
\end{align}
We show below in Sec.~\ref{SEChyperfunctionalEntropy} that the
hyperfluctuation correlation functions $\chi_\rho^A(\rv)$ and
$\chi_G^A(r)$ emerge naturally via an extended ensemble that is used
for the formulation of the entropic hyperfunctional theory, following
the lines of argumentation of Refs.~\cite{sammueller2024hyperDFT,
  sammueller2024whyhyperDFT, sammueller2024multihyperDFT}.

}

\section{Entropy functional forms}
\label{SECentropyFunctionalForms}

{\ms A localized version of the excess entropy functional is given in
  Sec.~\ref{SECentropyLocalizedEntrropyFunctional}. The reduced free
  energy constrained search formulation is developed in
  Sec.~\ref{SECreducedFreeEnergyPicture}.
The pairwise excess entropy approximation is shown to be equivalent to
the second-virial excess free energy approximation in
Sec.~\ref{SECpairEntropyApproximation}. Further simplistic
approximation of the excess entropy yields the standard mean-field
(random phase) free energy functional in Sec.~\ref{SECmeanField}.}

\subsection{Localized excess entropy density}
\label{SECentropyLocalizedEntrropyFunctional}

With a view towards constructing analytical or neural entropy
functionals we provide a re-formulation based on a localized excess
entropy density \cite{schmidt2011internalEnergy}. Without loss of
generality \cite{sammueller2024pairmatching} we consider the excess
entropy functional to be of the following spatially localized form:
\begin{align}
  S_\rmexc[\rho,\go] &=
  \int d\rv \rho(\rv) \sdensity(\rv;[\rho,\go]),
\end{align}
where the entropy density $\sdensity(\rv;[\rho,\go])$ is a
position-dependent functional of $\rho(\rv)$ and $\go(r)$. Then the
direct correlation functional \eqref{EQentropyDirectEntropic1APX}
attains the following form:
\begin{align}
  c_\rho(\rv;[\rho,\go]) &= \sdensity(\rv;[\rho,\go])
  + \int d\rv' \rho(\rv')
  \frac{\delta \sdensity(\rv';[\rho,\go])}{\delta\rho(\rv)}.
  \label{EQentropyDirectOne1Rewriting}
\end{align}
In the related excess free energy density functional context an
exchange of the arguments $\rv$ and $\rv'$ was used for practical
implementation \cite{sammueller2024pairmatching}. Furthermore the
functional derivative \eqref{EQentropyDirectEntropic2APX} becomes:
\begin{align}
  c_\go(\rv;[\rho,\go]) &=
  \int d\rv \rho(\rv) 
  \frac{\delta \sdensity(\rv;[\rho,\go])}{\delta \go(r)}.
  \label{EQentropyDirectOne2Rewriting}
\end{align}

The reformulation \eqref{EQentropyDirectOne1Rewriting} and
\eqref{EQentropyDirectOne2Rewriting} could help to develop machine
learning methods to train $\sdensity(\rv;[\rho,\go])$ as a neural
functional based on the Euler-Lagrange equations together with
corresponding simulation results for $\rho(\rv)$ and $\go(r)$ for
prescribed randomized forms of $\beta\mu$, $\beta V_\rmext(\rv)$,
and~$\beta\phi(r)$.

\subsection{Reduced free energy functional}
\label{SECreducedFreeEnergyPicture}
As an alternative yet equivalent formulation one may express the grand
potential metadensity functional
\eqref{EQentropyOmegaDecompositionAPX} as the following sum of the
reduced intrinsic free energy functional $F_\rmred[\rho,G]$, the
interparticle potential energy, the external potential energy, and the
chemical contribution:
\begin{align} 
  \beta \Omega[\rho,G] &= 
  \beta F_\rmred[\rho,G] 
  + \int_0^\infty dr G(r) \beta\phi(r)
  \notag\\&\quad 
  +\int d\rv   \rho(\rv) [\beta V_\rmext(\rv) - \beta \mu].
  \label{EQentropyGrandPotentialMetaDensityFunctional}
\end{align}
Here the reduced intrinsic free energy consists of the kinetic energy
and both (ideal and excess) entropic contributions,
\begin{align}
  F_\rmred[\rho,G] &= K[\rho] - TS[\rho,G],
  \label{EQentropyFredViaKTS}
\end{align}
and we recall the entropic splitting
\eqref{EQentropySidSexcSplittingAPX}. In contrast to the standard
density functional treatment \cite{hansen2013, evans1979, evans1992,
  evans2016}, the intrinsic potential energy is not included in
$F_\rmred[\rho,G]$, but rather accounted for separately via the second
term in \eqr{EQentropyGrandPotentialMetaDensityFunctional}. Hence, as
emphasized above, the interparticle potential is treated on an equal
footing with the external potential energy [third term in
  \eqr{EQentropyGrandPotentialMetaDensityFunctional}].

Comparison of the splitting
\eqref{EQentropyGrandPotentialMetaDensityFunctional} with Mermin's
functional~\eqref{EQentropyMerminFunctionalDefinitionAPX} then yields:
\begin{align} 
  \beta F_\rmred[\rho,G] &= \min_{f\to \rho,G} \Tr f
  \Big( \ln f +\beta\sumi\frac{\pv_i^2}{2m} \Big),
  \label{EQentropyFunctionalViaMaximizatoin}
\end{align}
which follows from analogous argumentation as presented in
Sec.~\ref{SECentropyAppendixOmegaDerivation}. One can then check
explicitly that the functionals for the reduced free energy
$F_\rmred[\rho,G]$, as given by
\eqr{EQentropyFunctionalViaMaximizatoin}, the kinetic energy
$K[\rho]$, as given via \eqr{EQentropyKineticEnergyFunctional}, and
the entropy $S[\rho,G]$, as given by
\eqr{EQentropyViaConstrainedMaximizationAPX}, satisfy the
relationship~\eqref{EQentropyFredViaKTS}.

\subsection{Virial approximation}
\label{SECpairEntropyApproximation}
We consider the following non-local approximation for the excess
entropy metadensity functional:
\begin{align}
  S_\rmexc[\rho,\go] &= \frac{k_B}{2}
  \int d\rv d\rv' \rho(\rv) \rho(\rv')
  \sbond(|\rv-\rv'|;[\go]).
  \label{EQentropySpairBondFormMain}
\end{align}
The convolution kernel $\sbond(r;[\go])$ is a (reduced) `entropic bond',
which carries nonlinear dependence on the global pair distribution
function $\go(r)$ via:
\begin{align}
  \sbond(r;[\go]) &= 
  -\go(r) \ln \go(r) + \go(r) - 1.
  \label{EQentropyBOFRMain}
\end{align}
In general $\sbond(r;[\go]) \leq 0$ with equality holding for the
ideal gas, $\go(r)=1$. We recall that $\go(r)=1$ implies
$G(r)=G_\rmid(r;[\rho])$ for any form of $\rho(\rv)$, as follows from
the definition \eqref{EQentropyGiddy}. {\ms We demonstrate at the end
  of this subsection that the formal structure
  \eqref{EQentropySpairBondFormMain} and \eqref{EQentropyBOFRMain} is
  consistent with the standard second-virial level approximation for
  the excess free energy density functional. Here we first lay out
  several consequences, as these serve to illustrate the general
  structure of the entropy functional approach.}

The pairwise approximation \eqref{EQentropySpairBondFormMain} can be
re-written equivalently as:
\begin{align}
  S_\rmexc[\rho,\go] &= 
  k_B \int_0^\infty dr  G_\rmid(r;[\rho])  \sbond(r;[\go]),
  \label{EQentropySpairDefinitionRewrittenMain}
\end{align}
which follows from inserting the definition \eqref{EQentropyGiddy} and
carrying out the pair distance integral. The re-writing
\eqref{EQentropySpairDefinitionRewrittenMain} highlights that the
approximation \eqref{EQentropySpairBondFormMain} is `distance-local'
with respect to its dependence on the pair distance variable~$r$.

A further alternative functional form is obtained from inserting the
definition \eqref{EQentropyBOFRMain} into the integral
\eqref{EQentropySpairDefinitionRewrittenMain} and expressing the
variable $\go(r)$ as the ratio \eqref{EQentropySmallgAsRatio}. This
yields
\begin{align}
  & S_\rmexc[\rho,G]=\notag\\&\;
  -k_B \int_0^\infty \!\!dr \Big[
    G(r) \ln\Big( \frac{G(r)}{G_\rmid(r;[\rho])}\Big)
    -G(r) + G_\rmid(r)\Big],
  \label{EQentropySpairDefinitionMain}
\end{align}
where the functional dependence is now on $\rho(\rv)$ and $G(r)$
rather than on $\rho(\rv)$ and $\go(r)$ as is the case in the forms
\eqref{EQentropySpairBondFormMain} and
\eqref{EQentropySpairDefinitionRewrittenMain}.

The approximative pair entropy functional
functional~\eqref{EQentropySpairDefinitionRewrittenMain} generates
entropic direct correlation functionals \eqref{EQentropyDirectcrhog}
and \eqref{EQdirectconeAlternative2} of the following forms:
\begin{align}
  c_{\rho|\go}(\rv;[\rho,\go]) &=
  (\sbond\star\rho)(\rv),
  \label{EQentropySpairDerivative1Main} \\
  c_\go(r;[\rho,\go]) &=
  -G_\rmid(r;[\rho])\ln \go(r),
  \label{EQentropySpairDerivative2Main}
\end{align}
where the star denotes the convolution as before, here of the entropic
bond~$\sbond(r;[\go])$ and the density profile $\rho(\rv)$,
\begin{align}
  (\sbond\star \rho)(\rv) & = 
  \int d\rv' \rho(\rv') \sbond(|\rv-\rv'|;[\go]).
  \label{EQentropyConvolutionBStarRho}
\end{align}

The alternative and equivalently formulation
\eqref{EQentropySpairDefinitionMain} generates via the `original'
formulation of the direct correlation functionals
\eqref{EQentropyDirectEntropic1APX} and
\eqref{EQentropyDirectEntropic2APX} the following forms:
\begin{align}
  c_{\rho}(\rv;[\rho,G]) &=
  (\sbond\star\rho)(\rv)
  -(\beta\phi \go\star\rho)(\rv)
  \\
  &= -(\psi \star \rho)(\rv)  
  \label{EQentropySpairDerivative3Main} \\
  c_G(r;[\rho,G]) &= -\ln \go(r),
  \label{EQentropySpairDerivative4Main}
\end{align}
where we have defined a scaled free energy bond,
\begin{align}
  \psi(r) &= \beta\phi(r)\go(r) -\sbond(r).
  \label{EQentropyPsiBond}
\end{align}
As a consistency check, the approximative forms
\eqref{EQentropySpairDerivative1Main} and
\eqref{EQentropySpairDerivative3Main} are consistent with the exact
direct correlation functional relationship
\eqref{EQentropyEntropicDirectCorrelationRelationAgain}, which we
repeat for convenience: $c_{\rho|\go}(\rv;[\rho,\go]) =
c_\rho(\rv;[\rho,G]) + (\beta\phi \go \star \rho)(\rv)$.

Using the result \eqref{EQentropySpairDerivative4Main} in the
Euler-Lagrange equation~\eqref{EQentropyEL2rhoGAPX} yields
$\go(r)=\e^{-\beta\phi(r)}$. Via insertion into the entropic bond
\eqref{EQentropyBOFRMain} the free energy bond
\eqref{EQentropyPsiBond} subsequently becomes $\psi(r)
=1-\e^{-\beta\phi(r)} = -f(r)$, where $f(r)$ is the standard
Mayer-bond \cite{hansen2013}. Inserting this result into the
corresponding form of the direct correlation functional
\eqref{EQentropySpairDerivative3Main} allows one to formulate the
Euler-Lagrange equation \eqref{EQentropyEL1rhoGAPX} in the following
approximate form:
\begin{align}
  \ln\rho(\rv) &= 
  (f \star \rho)(\rv) - \beta V_\rmext(\rv) + \beta \mu,
  \label{EQEulerLagrangeSecondVirial}
\end{align}
The integral equation \eqref{EQEulerLagrangeSecondVirial} is identical
in form to the Euler-Lagrange equation
\eqref{EQentropyEulerLagrangeStandardDFT} of standard density
functional theory, when using the low-density limit (second-virial
level) of the excess free energy functional, $\beta F_\rmexc[\rho] =
-\int d\rv d\rv'\rho(\rv)\rho(\rv') f(|\rv-\rv'|)/2$, which generates
via the functional derivative \eqref{EQc1Standard} the corresponding
one-body direct correlation functional $c_1(\rv;[\rho]) = \int
d\rv'\rho(\rv')f(|\rv-\rv'|) = (f \star \rho)(\rv)$.

{\ms
The general setup of the present entropy functional theory implies the
functional dependence on the specific arguments $\rho(\rv)$ and
$\go(r)$, see the approximative excess entropy functional
$S_\rmexc[\rho,\go]$, as is defined in the bilinear integral form
\eqref{EQentropySpairBondFormMain} with the entropic convolution bond
\eqref{EQentropyBOFRMain}. This functional dependence does not follow
in full from from merely applying the parametric temperature
derivative \eqref{EQentropySAsParametricDerivative} to the standard
second-virial free energy functional $F_\rmexc[\rho,\beta\phi]$. The
latter route rather yields the corresponding entropic metadensity
functional $S_\rmexc[\rho,\beta\phi]$ and we sketch the derivation. We
recall $F_\rmexc[\rho,\beta\phi] = -k_BT\int d\rv
d\rv'\rho(\rv)\rho(\rv') f(|\rv-\rv'|)/2$ and apply the general
mechanism \eqref{EQentropySAsParametricDerivative}:
$S_\rmexc[\rho,\beta\phi] = -\partial
F_\rmexc[\rho,\beta\phi]/\partial T$. Also recalling the form of the
Mayer bond as $f(r)=\e^{-\beta\phi(r)}-1$, the relevant temperature
derivative follows as $\partial [Tf(r)]/\partial T=f(r) + T\partial
f(r)/\partial T = f(r) + \beta\phi(r)\e^{-\beta\phi(r)}$. Collecting
the results gives the following standard form of the second-virial
excess entropy density functional:
\begin{align}
  S_\rmexc[\rho,\beta\phi]&=
  \frac{k_B}{2}\int d\rv d\rv' \rho(\rv) \rho(\rv') 
  \big(f +  \beta\phi\e^{-\beta\phi}\big),
  \label{EQsecondViralSexcMetaDensity}
\end{align}
where the arguments of $f(|\rv-\rv'|)$ and $\phi(|\rv-\rv'|)$ have
been suppressed in the integrand for clarity.  Hence the density
functional dependence on $\rho(\rv)$ is supplemented by the
metadensity dependence on $\beta\phi(r)$, which is explict in the
virial series. The result \eqref{EQsecondViralSexcMetaDensity} is
distinct from the universal excess entropy functional
$S_\rmexc[\rho,\go]$, which, while depending functionally on
$\rho(\rv)$, also features functional dependence on the global pair
distribution function $\go(r)$, see
Eqs.~\eqref{EQentropySpairBondFormMain} and~\eqref{EQentropyBOFRMain}.

To make the connection of the metadensity functional
\eqref{EQsecondViralSexcMetaDensity} with the universal excess entropy
functional $S_\rmexc[\rho,\go]$, one needs to realize that using the
low-density limit $\go(r) = \e^{-\beta\phi(r)}$ allows one to re-write
the convolution bond that appears in the functional
\eqref{EQsecondViralSexcMetaDensity} as $f(r) +
\beta\phi(r)\e^{-\beta\phi(r)} = -\go(r)\ln\go(r) + \go(r) - 1
=\sbond(r,[\go])$. This additional step yields consistency with the
definition of the entropic bond~\eqref{EQentropyBOFRMain} and thus
validates the non-local entropy
functional~\eqref{EQentropySpairBondFormMain} on the second-virial
level.

}

\subsection{Mean-field approximation}
\label{SECmeanField}

In view of the relationship
\eqref{EQentropyEntropicDirectCorrelationRelationAgain}, which we
repeat for convenience: $c_{\rho|\go}(\rv;[\rho,\go]) =
c_\rho(\rv;[\rho,G]) + (\beta\phi \go \star \rho)(\rv)$, it is
interesting to consider approximations.  Inserting the assumption
$\go(r)=1$ on the left hand side of the relation
\eqref{EQentropyEntropicDirectCorrelationRelationAgain} renders this
zero, as there are no excess correlations in the ideal gas,
$c_{\rho|\go}(\rv;[\rho,\go=1])=0$. Then re-arranging the right hand
side \eqref{EQentropyEntropicDirectCorrelationRelationAgain} yields
the approximation $c_\rho(\rv;[\rho,G])=-(\beta\phi\star
\rho)(\rv)$. Inserting this result on the left hand side of the
`position' Euler-Lagrange equation \eqref{EQentropyEL1rhoGAPX} yields
the following integral equation for the density profile:
\begin{align}
  \ln\rho(\rv) &= 
  -(\beta\phi\star\rho)(\rv)  - \beta V_\rmext(\rv) + \beta\mu.
  \label{EQentropyMeanFieldEL}
\end{align}

The form \eqref{EQentropyMeanFieldEL} is identical to the
Euler-Lagrange equation \eqref{EQentropyEulerLagrangeStandardDFT} of
standard classical density functional theory, when using the bilinear
mean-field free energy approximation, $\beta F_\rmexc[\rho,\beta\phi]
= -\int d\rv d\rv'\rho(\rv)\rho(\rv')\beta \phi(|\rv-\rv'|)/2$.

The present ease in deriving systematically the widely-used
approximation~\eqref{EQentropyMeanFieldEL} gives much confidence in
the general prowess of the entropic metadensity functional framework
for capturing the many-body physics.

{\ms

\section{Generalizations}
\label{SECgeneralization}

We next present several relevant conceptual generalizations of the
entropy functional framework.  We start by addressing general
many-body interparticle interaction potentials beyond the pairwise
contribution, first via parametric dependence in
Sec.~\ref{SECentropyManyBodyInteractions}. We then develop the
universal formal entropy functional dependence in
Sec.~\ref{SECviaReductionOperators}. This requires one to be able to
express the Hamiltonian via $n$-body reduction operators $\hat
G_n(R_n)$, which we introduce in generalization of the pair reduction
operator $\hat G(r)$. The approach delivers in principle significant
gain in efficiency over the explicit $n$-body density resolution, as
is argued in Sec.~\ref{SECmanyBodyPracticalities}. We use both
many-body methods to formulate an entropy hyperfunctional theory for
the description of the thermal behaviour of general observables in
Sec.~\ref{SEChyperfunctionalEntropy}.  We then return to pairwise
interacting system and describe the entropy functional framework for
multi-component classical systems in Sec.~\ref{SECmultiComponent} and
for quantum statistical mechanics in Sec.~\ref{SECquantumEntropy}.

}

\subsection{Many-body parametric dependence}
\label{SECentropyManyBodyInteractions}

In generalization of the pair form
\eqref{EQentropyInterparticlePotentialPairwiseAPX} of the
interparticle interaction potential, we consider Hamiltonians that
feature additional three- and higher-body terms. We hence take the
interparticle potential to possess the following generalized form:
\begin{align}
  \beta u(\rv^N) &= \frac{1}{2}\sum_{i=1}^N\sum_{j=1,j\neq i}^N 
  \beta\phi(|\rv_i-\rv_j|)\notag\\&\qquad
  + \beta \Delta u(\rv^N;\{u_\alphanew\}  ),
\end{align}
where $\Delta u(\rv^N;\{u_\alphanew\})$ constitutes the beyond-pair
multi-body potential energy that depends on a set of parameters and
possibly functions $\{u_\alphanew\}$, which are indexed
by~$\alphanew$.

We work with the reduced free energy formulation of
Sec.~\ref{SECreducedFreeEnergyPicture}.  Retaining the dependence on
$\{u_\alphanew\}$ yields in generalization of the constrained search
\eqref{EQentropyFunctionalViaMaximizatoin} the reduced intrinsic free
energy functional as the following constrained minimization:
\begin{align}
  &  \beta F_\rmred[\rho,G,\{u_\alphanew\}] \; = -\beta F_\rmid[\rho]
  \notag\\&\qquad
  +\min_{f\to\rho,G}\Big[
    \Tr f \Big(\ln f  
    + \beta \sumi\frac{\pv_i^2}{2m}
    + \beta \Delta u(\rv^N;\{u_\alphanew\}
    )\Big)\Big],
  \label{EQFredDefinition}
\end{align}
which contains entropic and multi-body potential contributions, but is
freed of the ideal and pair interaction contributions.

Due to subtracting the ideal gas term, the excess contribution retains
the dependence on the form of the many-body interaction
potential. Hence, upon adding the pair interaction energy, one obtains
the extended excess free energy functional as:
\begin{align}
   \beta {\cal F}_\rmexc[\rho,G,\beta\phi,\{u_\alphanew\} ] &= 
   \beta F_\rmred[\rho,G,\{u_\alphanew\}] 
   \notag\\&\quad
   +\int_0^\infty dr G(r)\beta\phi(r).
\end{align}
This setup mirrors the parametric dependence of the free-energy based
metadensity functional theory \cite{kampa2024meta,
  kampa2026pairmatching, kampa2026spherical}. We can relate to the
free energy metadensity functional point of view of
Refs.~\cite{kampa2024meta, kampa2026pairmatching, kampa2026spherical}
via partial minimization,
\begin{align}
  \beta F_\rmexc[\rho,\beta\phi,\{u_\alphanew\}]
  &= \min_G \beta {\cal F}_\rmexc[\rho,G,\beta\phi,\{u_\alphanew\}],
\end{align}
which is in generalization of the pairwise form
\eqref{EQFexcFromPartialMinimizationAPX} that is described in
Sec.~\ref{SECrelationshipWithStandardDFT}.

{\ms

The reduced free energy functional \eqref{EQFredDefinition} generates
first-order direct correlation functionals according to the following
functional derivatives:
\begin{align}
  c_\rho(\rv;[\rho,G,\{u_\alphanew\}]) &=
  -\frac{\delta \beta F_\rmred[\rho,G,\{u_\alphanew\}]}{\delta \rho(\rv)}
  \Big|_{G,\{u_\alphanew\}},
  \\
  c_G(r;[\rho,G,\{u_\alphanew\}]) &=
  -\frac{\delta \beta F_\rmred[\rho,G,\{u_\alphanew\}]}{\delta G(r)}
  \Big|_{\rho,\{u_\alphanew\}}.
\end{align}
The above direct correlation functionals feature in the following two
coupled Euler-Lagrange equations:
\begin{align}
  c_\rho(\rv;[\rho,G,\{u_\alphanew\}]) &=
  \ln\big(\rho(\rv)\Lambda^d\big) + \beta V_\rmext(\rv) - \beta\mu,
  \label{EQentropyEL1rhoParmetric} 
  \\
  c_G(r;[\rho,G,\{u_\alphanew\}]) &=
  \beta\phi(r),
  \label{EQentropyEL2rhoParmetric} 
\end{align}
which extend the pairwise versions \eqref{EQentropyEL1rhoGAPX} and
\eqref{EQentropyEL2rhoGAPX} by including the dependence of the direct
correlation functionals on the parameters $\{u_\alphanew\}$, see the
left hand sides \eqref{EQentropyEL1rhoParmetric} and
\eqref{EQentropyEL2rhoParmetric}.

We turn next to the alternative fully-resolved treatment of many-body
interaction potentials, as is based on the introduction of reduction
operators that generalize $\hat G(r)$ beyond the pairwise case.

}

{\ms

\subsection{Many-body reduction operators}
\label{SECviaReductionOperators}

Percus' stragegy to address entropy in systems that are governed by
many-body interparticle interactions entails having to resolve all
required three- and higher-body density distribution functions
explicitly~\cite{percus1994aspects}. His approach hence differs from
the present method of reducing both formal and computational
complexity by only resolving the {\it required} information to express
the interparticle interaction energy. On the pairwise level this
information is provided by~$G(r)$, we recall the definition
\eqref{EQentropyGhatAPX} of the corresponding operator.

The reduction strategy can be extended formally to interparticle
interaction potentials that contain $n$-body contributions.  As an
initial step, one is required to identify a set of appropriate
`reduction operators' $\hat G_\aalpha(R_\aalpha)$, where the argument
$R_\aalpha$ denotes a collection of e.g.\ distance, angular, or more
general geometric variables or measures that characterize
configurations of $n$ particles, in generalization of the dependence
of $\hat G(r)$ on pair distance~$r$.
The method is applicable to such forms of the interparticle
interaction potential $u(\rv^N)$ that one is able to express in the
following bilinear integral form:
\begin{align}
  u(\rv^N) &= 
  \sum_{\aalpha\in\calU} \int dR_\aalpha
  \hat G_\aalpha(R_\aalpha) \phi_\aalpha(R_\aalpha).
  \label{EQurnManyBody}
\end{align}
The integral $\int dR_n$ is over the respective (geometric) ranges of
the underlying individual parameters that constitute
each~$R_\aalpha$. The sum is over an index set $\calU$ that consists
of those orders $n$ that are required to express $u(\rv^N)$.  For
simultaneous pairwise and triplet interparticle interactions
$\calU=\{2,3\}$. Hence in typical cases the sum \eqref{EQurnManyBody}
will not be an infinite series. The pairwise term is contained for the
special case $\aalpha=2$ as $\hat G_{2}(R_{2})=\hat G(r)$,
$\phi_2(R_2)=\phi(r)$, and $R_2=r$; we turn to $n=3$ as a further
relevant special case below.  

In the notation the dependence of $\hat G_\aalpha(R_\aalpha)$ on the
microstate~$\rv^N$ is suppressed and indicated by the
caret. Crucially, the generalized potential functions
$\phi_\aalpha(R_\aalpha)$ depend merely on~$R_\aalpha$ but not
on~$\rv^N$, which is inline with the structure of the pairwise term,
$\phi(r)$.  By construction, functional differentiation of the
bilinear integrals \eqref{EQurnManyBody} retrieves the individual
reduction operators: $\hat G_\aalpha(R_\aalpha) = \delta
u(\rv^N)/\delta \phi_\aalpha(R_\aalpha)$.  The product notation $\hat
G_\aalpha(R_\aalpha) \phi_\aalpha(R_\aalpha)$ implies appropriate
contraction of any potentially underlying tensorial structure, such as
$ \hat G_\aalpha(R_\aalpha):\phi_\aalpha(R_\aalpha)$ in case $\hat
G_\aalpha(R_\aalpha)$ and $\phi_\aalpha(R_\aalpha)$ are matrix-valued,
such that the result is the (scalar) potential
energy~\eqref{EQurnManyBody}.

Based on the integral form \eqref{EQurnManyBody}, the entropy
functional construction follows through in straightforward extension
over the pairwise case.  One imposes the following set of joint
constraints:
\begin{align}
  \rho(\rv) &= \Tr f \hat\rho(\rv), 
  \label{EQGextendedConstraint1}  \\
  G_\aalpha(R_\aalpha)&=\Tr f \hat  G_\aalpha(R_\aalpha), 
  \label{EQGextendedConstraint2}
\end{align}
in generalization of the density \eqref{EQentropyConstraint1APX} and
pair distance distribution constraints
\eqref{EQentropyConstraint2APX}. We refer to the $G_n(R_n)$ as the
($n$-body) `reduction fields'.

The Levy search for the minimum of Mermin's functional under the
simultaneous set of constraints \eqref{EQGextendedConstraint1} and
\eqref{EQGextendedConstraint2} for all $n\in\calU$ then yields an
extended entropy functional $S[\rho,\{\nG\}]$, where $\{G_\aalpha\}$
denotes the
set of all reduction fields $G_\aalpha(R_\aalpha)$; we recall that
this will include, in general, the global pair reduction function
$G(r)$ as a special case \eqref{EQentropyGhatAPX}.  The coupled
Euler-Lagrange equations follow as:
\begin{align}
  c_{\rho}(\rv;[\rho,\{\nG\}]) &= \ln\big(\rho(\rv)\Lambda^d\big)
  + \beta V_\rmext(\rv) - \beta\mu,
  \label{EQmultiBodyEL1explicit}\\
  c_{\aalpha}(R_\aalpha;[\rho,\{\nG\}]) &= 
  \beta \phi_\aalpha(R_\aalpha),
  \label{EQmultiBodyEL2explicit} 
\end{align}
which generalize the Euler-Lagrange equations
\eqref{EQentropyEL1rhoGAPX} and \eqref{EQentropyEL2rhoGAPX}. The
generalized entropic direct correlation functionals
$c_{\rho}(\rv;[\rho,\{\nG\}])$ and
$c_{\aalpha}(R_\aalpha;[\rho,\{\nG\}])$ are defined as functional
derivatives of the excess entropy functional $S_\rmexc[\rho,\{\nG\}]$
via:
\begin{align}
  c_{\rho}(\rv;[\rho,\{\nG\}]) &= 
  \frac{\delta  S_\rmexc[\rho,\{\nG\}]/k_B}{\delta \rho(\rv)}
  \Big|_{\{\nG\}},
  \label{EQextendedDirectCorrelationFunctionMultiBody1}\\
  c_{\aalpha}(R_\aalpha;[\rho,\{\nG\}]) &= \frac{\delta
    S_\rmexc[\rho,\{\nG\}]/k_B} {\delta G_\aalpha(R_\aalpha)}
  \Big|_{\rho,\{G_{\breve\aalpha}\}},
  \label{EQextendedDirectCorrelationFunctionMultiBody2}
\end{align}
in generalization of the
definitions~\eqref{EQentropyDirectEntropic1APX} and
\eqref{EQentropyDirectEntropic2APX}. We have used the breve shorthand
notation $\{G_{\breve\aalpha}\} = \{G_{\aalpha'}| \aalpha'\in\calU,
\aalpha'\neq\aalpha\}$ to exclude the specific element $n$ with
respect to which the functional derivative
\eqref{EQextendedDirectCorrelationFunctionMultiBody2} is taken. The
excess entropy functional arises from the splitting of the total
entropy functional into ideal and excess contributions,
$S[\rho,\{\nG\}] = S_\rmid[\rho] + S_\rmexc[\rho,\{\nG\}]$, with the
ideal gas entropy functional~\eqref{EQentropySidOneDefinition} and in
generalization of the ideal-excess
splitting~\eqref{EQentropySidSexcSplittingAPX} for pairwise-only
interacting systems.

The mean energy functional $E[\rho,\{\nG\}]$ inherits the many-body
interaction terms from the structure of the bilinear operator
form~\eqref{EQurnManyBody} and it is hence explicitly given by:
\begin{align}
  E[\rho,\{\nG\}] &= K[\rho]
  + \sum_{\aalpha\in\calU} 
  \int dR_\aalpha G_\aalpha(R_\aalpha) \phi_\aalpha(R_\aalpha)
  \notag\\&\quad
  + \int d\rv \rho(\rv) V_\rmext(\rv),
  \label{EQextendedMeanEnergyMultiBody}
\end{align}
which is in generalization of the pairwise form
\eqref{EQmeanEnergyFunctionalAPX}.  Analogously, the grand potential
functional \eqref{EQmetaentropyOmegaViaEnergyEntropyAPX} becomes:
\begin{align}
  \Omega[\rho,\{\nG\}] 
  &= E[\rho,\{\nG\}] -
  TS[\rho,\{\nG\}] - \mu N[\rho],
  \label{EQextendedGrandPotentialMultiBody}
\end{align}
where $N[\rho]$ remains the trivial mean particle number
functional~\eqref{EQmeanParticleNumberDensityFunctional}.  For
completeness, the explicit Euler-Lagrange equations
\eqref{EQmultiBodyEL1explicit} and \eqref{EQmultiBodyEL2explicit}
arise respectively from the joint minimization conditions:
\begin{align}
  \frac{\delta \Omega[\rho,\{\nG\}]}{\delta \rho(\rv)}
  \Big|_{\{\nG\}}&= 0
  \qquad \textrm{(min)},
  \label{EQmultiBodyEL1compact} \\
  \frac{\delta \Omega[\rho,\{\nG\}]}{\delta \nG(R_\aalpha)} 
  \Big|_{\rho,\{\G_{\breve\aalpha}\}}&= 0
  \qquad \textrm{(min)},
  \label{EQmultiBodyEL2compact}
\end{align}
which generalize the joint minimization conditions
\eqref{EQentropyEL1rawAPX} and \eqref{EQentropyEL2rawAPX} of the
pairwise case.

In generalization of the definitions
\eqref{EQentropySecondOrderDirect1}--\eqref{EQentropySecondOrderDirect4},
we obtain second-order direct correlation functionals as functional
derivatives of the first-order direct correlation functionals
\eqref{EQextendedDirectCorrelationFunctionMultiBody1} and
\eqref{EQextendedDirectCorrelationFunctionMultiBody2} in the following
forms:
\begin{align}
  c_{\rho\rho}(\rv,\rv';[\rho,\{\nG\}]) &=
  \frac{\delta c_\rho(\rv;[\rho,\{\nG\}])}{\delta \rho(\rv')}
  \Big|_{\{\nG\}}, 
  \label{EQmultiBodySecondOrderDirect1}\\
  c_{\rho \aalpha'}(\rv,R'_{\aalpha'};[\rho,\{\nG\}]) &=
  \frac{\delta c_\rho(\rv;[\rho,\{\nG\}])}{\delta G_{\aalpha'}(R'_{\aalpha'})}
  \Big|_{\rho,\{\G_{\breve\aalpha}\}},
  \\
  c_{\aalpha\rho}(R_\aalpha,\rv';[\rho,\{\nG\}]) &=
  \frac{\delta c_\aalpha(R_\aalpha;[\rho,\{\nG\}])}{\delta\rho(\rv')}
  \Big|_{\{\nG\}},\\
  c_{\aalpha\aalpha}(R_\aalpha,R'_{\aalpha'};[\rho,\{\nG\}]) &=
  \frac{\delta c_\aalpha(R_\aalpha;[\rho,\{\nG\}])}{\delta G_{\aalpha'}(R'_{\aalpha'})}
  \Big|_{\rho,\{\G_{\breve\aalpha}\}}.
  \label{EQmultiBodySecondOrderDirect4}
\end{align}

The second-order direct correlation functionals
\eqref{EQmultiBodySecondOrderDirect1}--\eqref{EQmultiBodySecondOrderDirect4}
feature in coupled entropic Ornstein-Zernike equations, which follow
from corresponding generalization of the pairwise system of equations
\eqref{EQentropyOZ1}--\eqref{EQentropyOZ4}. They are also important in
the hyperfunctional theory for the statistical mechanical behaviour of
general observables, as is laid out below in
Sec.~\ref{SEChyperfunctionalEntropy}, after we first describe some
matters of potential practical use.

\subsection{Many-body reduction efficiency}
\label{SECmanyBodyPracticalities}

The functional formulation of Sec.~\ref{SECviaReductionOperators}
offers a formal framework for achieving potentially dramatic reduction
in computational complexity over Percus' $n$-body
method~\cite{percus1994aspects}. We recall that his method requires
one, in principle, to work explicitly with joint position dependencies
on $\rv,\rv',\rv'',...$, in contrast to the present reduction to the
set $\{R_\aalpha\}$.  Feasibility is not merely an academic issue,
given the significant interest in the study of many-body potentials,
such as monatomic water~\cite{coe2022prl, coe2023} and machine-learned
interparticle potentials (MLIPs) that emerge from an underlying {\it
  ab initio} treatment. In pioneering work, these MLIPs were already
demonstrated successfully to provide a sound basis for neural
functional learning \cite{bui2026abinitio}.  For an enlightening
discussion of important conceptual aspects that arise from the study
of many-body interactions, we refer the Reader to the recent
Perspective by Villegas and Melon~\cite{villegas2026}.

It is hence interesting to relate the complexity of the present
approach to Percus' $n$-body expansion, focusing on $d=3$ for
concreteness and assuming translational and rotational equivariance of
$u(\rv^N)$. As emphasized throughout, the two-body term replaces the
dependence on the six variables $\rv,\rv'$ by the dependence on the
single parameter $r=|\rv-\rv'|$, as it arises from the three
translational and two rotational space invariances. Proceeding to
$n=3$, any triplet of particles is determined by three variables, such
as the lengths $r,r',r''$ of the triangle sides. Hence in contrast to
resolving the nine degrees of freedom $\rv,\rv',\rv''$ of the
three-body density distribution, merely three (scalar) parameters are
involved. In $d=3$ this is on par with fully inhomogeneous position
dependence of~$\rho(\rv)$. To be explicit, a suitable form of the
three-body reduction operator is
\begin{align}
  \hat G_3(r,r',r'')
  &=\frac{1}{6}\sum_{ijk(\neq)} \delta(r-|\rv_i-\rv_j|)
  \delta(r'-|\rv_j-\rv_k|)\notag\\
  &\qquad\quad\quad
  \times\delta(r''-|\rv_k-\rv_i|),
\end{align}
where the sum is over all distinct triplets, $\sum_{ijk(\neq)} =
\sum_{i=1}^N\sum_{j=1,j\neq i}^N\sum_{k=1,k\neq i,j}^N$. The
corresponding integration domain is then given by the three distance
integrals $\int dR_3 = \int_0^\infty dr \int_0^\infty dr'
\int_0^\infty dr''$.

\smallskip

The above favourable equivariant reduction is exhausted though when
considering four- and higher-body terms. The full resolution acquires
three additional variables at each increase in $n$. However, benefits
continue to be significant in case that these four- and higher-body
terms depend on merely reduced (geometric) measures of the respective
$n$-particle configurations, as can be encapsulated by careful
construction of the reduction operators $\hat G_\aalpha(R_\aalpha)$,
provided that this is feasible for the given Hamiltonian.  We
emphasize that the dependence of $\hat G_\aalpha(R_\aalpha)$ on the
particle configuration~$\rv^N$ can thereby be of an algorithmic nature
and thus include operations such as clustering, sorting, and binning;
see Refs.~\cite{sammueller2024hyperDFT, sammueller2024whyhyperDFT,
  sammueller2024multihyperDFT} for corresponding neural hyperdensity
functional work.

\subsection{Entropic hyperfunctional theory}
\label{SEChyperfunctionalEntropy}

Hyperdensity functional theory \cite{sammueller2024hyperDFT,
  sammueller2024whyhyperDFT, sammueller2024multihyperDFT} is a recent
variational approach for the thermal behaviour of general observables
$\hat A$, as we describe briefly in the following to provide some
context.  The standard Mermin-Evans functional map \cite{mermin1965,
  evans1979} ascertains that for given interparticle interaction
potential $u(\rv^N)$, the thermal mean of any observable $\hat A$
constitutes a universal density functional $A[\rho]$, in the sense
that this is independent of the external potential.  Hyperdensity
functional theory \cite{sammueller2024hyperDFT,
  sammueller2024whyhyperDFT, sammueller2024multihyperDFT} turns this
formal relationship into a concrete computational scheme. The relevent
direct and hyperdirect correlation functionals were shown to be
amenable to local functional learning, based on simulation data. The
theory gives access to local fluctuation profiles $\chi_A(\rv)$ that
characterize the correlation of the local density and the value of the
`hyperobservable' across the entire thermal ensemble, see the
covariance expression \eqref{EQchiAdensity} on which we extend
below. A spatially localized hyperdirect correlation functional
$c_A(\rv;[\rho])$ plays a similar role as the standard one-body direct
correlation functional $c_1(\rv;[\rho])$ in providing functional
closure on the one-body level. The interrelationship is provided by an
exact hyper-Ornstein-Zernike equation \cite{sammueller2024hyperDFT,
  sammueller2024whyhyperDFT, sammueller2024multihyperDFT}:
\begin{align}
  c_A(\rv;[\rho]) &= \frac{\chi_A(\rv)}{\rho(\rv)}
  - \int d\rv' c_2(\rv,\rv';[\rho]) \chi_A(\rv'),
\end{align}
where $c_2(\rv,\rv';[\rho])=\delta c_1(\rv;[\rho])/\delta\rho(\rv')$
is the standard two-body direct correlation functional
\cite{evans1979, hansen2013}. The mean hyperdensity functional
$A[\rho]$ generates the first-order hyperdirect correlation functional
via functional differentiation:
\begin{align}
  c_A(\rv;[\rho]) &= \frac{\delta A[\rho]}{\delta \rho(\rv)}.
  \label{EQhyperDFTcA}
\end{align}
The functional derivative \eqref{EQhyperDFTcA} can be efficiently
inverted via functional line integration to provide $A[\rho]$ once
$c_A(\rv;[\rho])$ is represented as a neural network
\cite{sammueller2024hyperDFT, sammueller2024whyhyperDFT,
  sammueller2024multihyperDFT}.

Here we provide the entropically resolved version of the hyperdensity
functional theory. We use both the parametric dependence framework of
Sec.~\ref{SECentropyManyBodyInteractions} and the reduction operator
formalism of Sec.~\ref{SECviaReductionOperators}. The definition of an
extended ensemble \cite{sammueller2024hyperDFT,
  sammueller2024whyhyperDFT, sammueller2024multihyperDFT} is based on
the extended Hamiltonian $H_A = H - \lambda \hat A/\beta$, where
$\lambda$ is a coupling parameter.  Building parametric
$\lambda$-derivatives then generates the relevant relationships.
Setting $\lambda=0$ after the derivatives have been taken restores the
thermal averages of the original ensemble
\cite{sammueller2024hyperDFT, sammueller2024whyhyperDFT,
  sammueller2024multihyperDFT}. We imply this convention throughout in
the following and suppress the dependence on $\lambda$ in the notation
for ease of reading.

{\ms

Based on the reduction operator form \eqref{EQurnManyBody} of the
Hamiltonian, we express the given hyperobervable $\hat A$ as the
following sum:
\begin{align}
  \hat A &= \hat A_\rmred +
  \sum_{n\in\calU} \int d R_\aalpha 
  \hat G_\aalpha(R_\aalpha)
  \psi_\aalpha(R_\aalpha),
  \label{EQAhatManyBody}
\end{align}
where the reduced hyperobervable $\hat A_\rmred$ is the remainder over
all terms that are accounted for by the sum over $n$, which contains
those orders~$n\in\calU$ that contribute to the Hamiltonian $H$. We
recall that the reduction operators $\hat G_n(R_n)$ carry the
dependence on the microstate $\rv^N$ whilst the functions
$\psi_\aalpha(R_\aalpha)$ depend only on the geometric parameters
$R_n$ and are specific to the form of $\hat A$.  We imply
$\psi_n(R_n)=0$ for those orders~$n$ that do not contribute to $\hat
A$.  The thus reduced operator $\hat A_\rmred$ carries all remaining
terms that are not expressible via the given reduction operators $\hat
G_n(R_n)$.

The framework allows one to consider forms of the hyperoberservable
$\hat A$ that may be `more complex' than the interparticle interaction
potential $u(\rv^N)$, such that $\hat A$ contains $n$-body terms that
do not feature in the Hamiltonian itself.  We refer to
Refs.~\cite{sammueller2024hyperDFT, sammueller2024whyhyperDFT,
  sammueller2024multihyperDFT} for the investigation of cluster
statistics in simple fluids as an example. Our following treatment is
general.

We recall $H_A = H - \lambda \hat A/\beta$, and define a
correspondingly extended interparticle interaction potential
$u_A(\rv^N) = u(\rv^N) - \lambda \hat A/\beta$. Generalized $n$-body
potentials follow as $\phi_n^A(R_n) = \phi_n(R_n) - \lambda
\psi_n(R_n)/\beta$. The extended interparticle potential can then be
expressed as
\begin{align}
  u_A(\rv^N) &=
  -\frac{\lambda}{\beta}\hat A_\rmred
  +\sum_{n\in \calU} \int dR_n \hat G_n(R_n) \phi^A_n(R_n).
\end{align}

In the thus extended ensemble \cite{sammueller2024hyperDFT,
  sammueller2024whyhyperDFT, sammueller2024multihyperDFT} one finds
the hyperfunctional as the parametric derivative $A[\rho,\{\nG\}] =
-\beta\partial \Omega[\rho,\{\nG\}]/\partial \lambda$.  Performing the
derivative, upon taking account of the minimum conditions
\eqref{EQmultiBodyEL1compact} and \eqref{EQmultiBodyEL2compact} while
observing that $\beta\partial\phi_n^A(R_n)/\partial \lambda =
\psi_n(R_n)$, yields the following functional identity:
\begin{align}
  A[\rho,\{\nG\}]  &= 
  A_\rmred[\rho,\{\nG\}] + 
  \sum_{n\in \calU} \int dR_n G_n(R_n)\psi_n(R_n).
  \label{EQAhyperfunctionalDefinition}
\end{align}
Thereby the reduced hyperfunctional $A_\rmred[\rho,\{\nG\}]$ contains
the effects that are not accounted for by the explicit sum over
$\aalpha$.  The reduced hyperfunctional is given specifically as the
parametric $\lambda$-derivative of the reduced free energy functional:
\begin{align}
  A_\rmred[\rho,\{\nG\}] &=
  -\frac{\partial \beta F_\rmred[\rho,\{\nG\}]}{\partial\lambda}
  \Big|_{\rho,\{\G_n\}},
  \label{EQAexcDefinition}
\end{align}
where we recall that $\lambda=0$ is set after the $\lambda$-derivative
has been taken, as we imply throughout \cite{sammueller2024hyperDFT,
  sammueller2024whyhyperDFT, sammueller2024multihyperDFT}.  The
reduced free energy functional $\beta F_\rmred[\rho,\{\nG\}]$, see
Sec.~\ref{SECentropyManyBodyInteractions} for the pairwise
formulation, inherits the dependence on $\lambda$ and it depends
functionally on $\rho(\rv)$ and on all $G_\aalpha(R_n)$ with
$n\in\calU$, as is consistent with the reduction operator formulation
of Sec.~\ref{SECviaReductionOperators}.

We lay out several consequences of the hyperfunctional theory by first
formulating entropic versions of locally resolved hyperfluctuation
profiles and hyperdirect correlation functionals
\cite{sammueller2024hyperDFT, sammueller2024whyhyperDFT,
  sammueller2024multihyperDFT}. We start by extending and generalizing
the prior definitions~\eqref{EQchiAdensity} and~\eqref{EQchiAGhat} by
the following set of hyperfluctuation profiles:
\begin{align}
  \chi^A_{\rho}(\rv) &= \cov(\hat A, \hat\rho(\rv)) 
  = \frac{\partial\rho(\rv)}{\partial\lambda },
  \label{EQchiAdensityFull}\\
  \chi^A_{\aalpha}(R_\aalpha) &= \cov(\hat A, \hat G_\aalpha(R_\aalpha))
  = \frac{\partial G_\aalpha(R_\aalpha)}{\partial \lambda},
  \label{EQchinMultiBody}
\end{align}
where the $\lambda$-derivatives are taken with constant $\mu$, $T$ and
with fixed potentials $V_\rmext(\rv)$ and $u(\rv^N)$.  The $n$-body
form \eqref{EQchinMultiBody} reduces for $n=2$ to the pair distance
hyperfluctuation profile $\chi^A_G(r) = \cov(\hat A, \hat G(r)) =
\partial G(r)/\partial \lambda$, see its definition
\eqref{EQchiAGhat}.

First-order hyperdirect correlation functionals are obtained via
functional differentiation of the reduced hyperfunctional
\eqref{EQAexcDefinition} as
\begin{align}
  c^A_\rho(\rv;[\rho,\{\nG\}]) &=
  \frac{\delta A_{\rmred}[\rho,\{\nG\}]}{\delta \rho(\rv)}
  \Big|_{\{\nG\}},\\
  c^A_{\aalpha}(R_\aalpha;[\rho,\{\nG\}]) &=
  \frac{\delta A_{\rmred}[\rho,\{\nG\}]}{\delta G_\aalpha(R_\aalpha)}
  \Big|_{\rho,\{\G_{\breve\aalpha}\}}.
\end{align}

Parametric $\lambda$-differentation of the Euler-Lagrange equations
\eqref{EQmultiBodyEL1explicit} and \eqref{EQmultiBodyEL2explicit}
yields the following set of exact entropic hyper-Ornstein-Zernike
equations:
\begin{align}
  & c^A_{\rho}(\rv;[\rho,\{\nG\}]) = \notag\\
  &\qquad \frac{\chi^A_{\rho}(\rv)}{\rho(\rv)} 
  - \int d\rv' c_{\rho\rho}(\rv,\rv';[\rho,\{\nG\}]) \chi^A_{\rho}(\rv')
  \notag  \\
  &\qquad -\sum_{\aalpha'\in\calU}\int dR'_{\aalpha'}
  c_{\rho \aalpha'}(\rv,R'_{\aalpha'};[\rho,\{\nG\}])
  \chi^A_{{\aalpha'}}(R'_{\aalpha'}),
  \\
  & c^A_{\aalpha}(R_\aalpha;[\rho,\{\nG\}]) =  \notag\\
  &\qquad  - \psi_\aalpha(R_\aalpha)
  -\int d\rv' c_{\aalpha\rho}(R_\aalpha,\rv';[\rho,\{\nG\}]) 
  \chi^A_{\rho}(\rv')
  \notag \\
  &\qquad - \sum_{\aalpha'\in\calU}
  \int dR'_{\aalpha '}
  c_{\aalpha {\aalpha'}}(R_\aalpha,R'_{\aalpha'};[\rho,\{\nG\}]) 
  \chi^A_{{\aalpha'}}(R'_{\aalpha'}),
\end{align}
where the second-order direct correlation functionals are given as the
excess entropy functional derivatives
\eqref{EQmultiBodySecondOrderDirect1}--\eqref{EQmultiBodySecondOrderDirect4}.

The present formalism provides a theoretical structure for carrying
out deep investigations into the behaviour of complex observables in
complex systems. The splitting of the hyperfunctional into reduced and
explict contributions, see the first and second term in the definition
\eqref{EQAhyperfunctionalDefinition}, mirrors the decomposition of
free energy into entropic and energetic constituents. The entropic
universality remains intact, as the hyperfunctional
$A[\rho,\{G_\aalpha\}]$ applies across all Hamiltonians with varying
forms of the interparticle potential functions $\phi_n(R_n)$ within
the considered orders $n\in\calU$. The formalism can help to address
the challenges in constructing first-principles-based machine-learning
schemes, using training data from many-body simulations, as we
re-iterate was demonstrated successfully for the free energy based
hyperdensity functional theory
\cite{sammueller2024hyperDFT,sammueller2024whyhyperDFT,
  sammueller2024multihyperDFT}.

}
}

{\ms

\subsection{Classical multi-component systems}
\label{SECmultiComponent}

We return to pairwise interparticle forces and address classical
mixtures with $M$ components and label the different species by
$\alpha=1,\ldots, M$.  Index sets ${\cal N}_\alpha$, which encompass
the particle indices $i$ of species $\alpha$, are used for the
bookkeeping.  The summation over the particles of species $\alpha$ can
then be written succinctly as $\sum_{i\in {\cal N}_\alpha}$.
We consider pairwise interacting systems with species-resolved pair
potentials $\phi_{\aas}(r)$ that act between species
$\alpha,\alpha'=1,\ldots,M$.  Species exchange symmetry is implied,
$\phi_\aas(r)=\phi_{\alpha'\alpha}(r)$.
The mixture is under the influence of species-resolved external
potentials $V_{\rmext,\alpha}(\rv)$. The thermodynamic parameters are
the species-resolved chemical potentials $\mu_\alpha$ and absolute
temperature $T$. We keep $N=\sum_{\alpha=1}^M N_\alpha$ as the total
number of particles, where $N_\alpha$ denotes the number of particles
of species $\alpha$.

We define partial density and pair reduction operators, in respective
generalization of the one-component versions
\eqref{EQdensityOperatorAPX} and \eqref{EQentropyGhatAPX}, as
\begin{align}
  \hat\rho_\alpha(\rv) &= \sum_{i \in \Nalpha} \delta(\rv-\rv_i),
  \label{EQdensityOperatorMixture}\\
  \hat G_\aas(r) &= \frac{1}{2}\sum_{i\in\Nalpha} \sum_{j\in\Nalphas,j\neq i}
  \delta(r-|\rv_i-\rv_j|)
  \label{EQGhatOperatorMixture}.
\end{align}
The Hamiltonian for the multi-species system can then be expressed as
\begin{align}
  H_{\!\cal N}  
  &= \sum_{\alpha=1}^M \sum_{i\in\Nalpha} \frac{\pv_i^2}{2m_\alpha}
  + \sum_{\alpha=1}^M\sum_{\alpha'=1}^M
  \int dr \hat G_\aas(r) \phi_\aas(r)
  \notag\\&\quad
  +\sum_{\alpha=1}^M
  \int d\rv\hat\rho_\alpha(\rv) V_{\rmext,\alpha}(\rv),
\end{align}
where the subscript indicates the shorthand ${\cal
  N}=\{\calN_\alpha\}=\{\calN_1, \calN_2, \ldots, \calN_M\}$ for all
index sets, in generalization of the dependence of $H_N$ on the number
of particle $N$ in the one-component system
\eqref{EQentropyHamiltonianRewrittenAPX}, and $m_\alpha$ is the
particle mass of species $\alpha$. As in the one-component system, the
operators \eqref{EQdensityOperatorMixture} and
\eqref{EQGhatOperatorMixture} are recovered by functional
differentiation with respect to the respective conjugate field: $\hat
\rho_\alpha(\rv) = \delta H_{\!\calN}/\delta V_{\rmext,\alpha}(\rv)$
and $\hat G_\aas(r) = \delta H_{\!\calN} / \delta \phi_\aas(r)$.

To formulate the statistical mechanics, we start with the standard
multi-component version of the classical trace:
\begin{align}
  \Trmix &= 
  \sum_{N_1=0}^\infty \sum_{N_2=0}^\infty \ldots  \sum_{N_M=0}^\infty
  \frac{h^{-d(N_1+N_2+\ldots N_M)}}{N_1!N_2!\ldots N_M!}
  \int  d\rv^N d\pv^N,
\end{align}
with grand partition sum $\Xi = \Trmix \e^{-\beta (H_\calN -
  \sum_{\alpha=1}^M \mu_\alpha N_\alpha)}$, equilibrium measure
$f_0=\e^{-\beta (H_\calN - \sum_{\alpha=1}^M \mu_\alpha
  N_\alpha)}/\Xi$, and grand potential
$\Omega_0(T,\{\mu\})=-k_BT\ln\Xi$, where $\{\mu\}$ denotes the set of
all chemical potentials $\mu_\alpha$.
The generalization of Mermin's functional
\eqref{EQentropyMerminFunctionalDefinitionAPX} to mixtures is
\begin{align}
  \beta \Omega_M[f] &=
  \Trmix f \Big(\ln f + \beta H_{\!\calN}
  - \beta\sum_{\alpha=1}^M \mu_\alpha  N_\alpha \Big),
\end{align}
where the trial probability distribution function $f(\rv^\calN,
\pv^\calN)$ depends on all phase space variables and all index sets
$\calN$.

Species-resolved sets of simultaneous constraints are provided via
generalization of the one-component versions
\eqref{EQentropyConstraint1APX} and \eqref{EQentropyConstraint2APX}
as:
\begin{align}
  \rho_\alpha(\rv) &= \Trmix f \hat\rho_\alpha(\rv),
  \label{EQentropyConstraint1Mixture}\\
  G_\aas(r) &= \Trmix f \hat G_\aas(r).
  \label{EQentropyConstraint2Mixture}
\end{align}
The constrained search then endows the grand potential with functional
dependence on sets of species-resolved distribution functions, for
which we use the shorthand notations
$\{\rho\}=\{\rho_1,\rho_2,\ldots,\rho_M\}$ and $\{{\sf G}\}=\{G_\aas;
\alpha,\alpha'=1,\ldots,M\}$. The constrained minimization yields
\begin{align}
  \Omega[\{\rho\},\{\sfG\}] &= 
  \min_{f\to\{\rho\},\{\sfG\}} \Omega_M[f],
  \label{EQentropyOmegaFunctionalFromMerminAPXMixture}
\end{align}
where $f\to\{\rho\},\{\sfG\}$ indicates the set of simultaneous
constraints \eqref{EQentropyConstraint1Mixture} and
\eqref{EQentropyConstraint2Mixture} for all $\alpha,
\alpha'=1,\ldots,M$.

The grand potential functional
\eqref{EQentropyOmegaFunctionalFromMerminAPXMixture} for the mixture
can be decomposed, in generalization of the one-component version
\eqref{EQmeanEnergyFunctionalAPX}, as
\begin{align}
  \Omega[\{\rho\},\{\sfG\}] &=
  E[\{\rho\},\{\sfG\}] - T S[\{\rho\},\{\sfG\}] 
  \notag\\&\quad
  - \sum_{\alpha=1}^M \mu_\alpha N[\rho_\alpha],
\end{align}
where $N[\rho_\alpha]= \int d\rv \rho_\alpha(\rv)$ is the trivial
global particle number density
functional~\eqref{EQmeanParticleNumberDensityFunctional} now acting on
each partial density profile. The multi-component mean energy
functional generalizes the one-component version
\eqref{EQmeanEnergyFunctionalAPX} in the following explicit form:
\begin{align}
  E[\{\rho\},\{\sfG\}] &=
  \sum_{\alpha=1}^M K[\rho_\alpha]
  \notag\\&\quad
  + \sum_{\alpha=1}^M \sum_{\alpha'=1}^M 
  \int_0^\infty dr G_\aas(r) \phi_\aas(r) 
  \notag\\&\quad
  +\sum_{\alpha=1}^M \int d\rv \rho_\alpha(\rv) V_{\rmext,\alpha}(\rv).
\end{align}
where the classical mean kinetic energy functional
\eqref{EQentropyKineticEnergyFunctional} remains given as
$K[\rho_\alpha]=(k_BT d/2)\int d\rv\rho_\alpha(\rv)$.

The multi-component entropy functional $S[\{\rho\},\{\sfG\}]$ is given
formally by the maximization under the joint species-resolved
constraints \eqref{EQentropyConstraint1Mixture} and
\eqref{EQentropyConstraint2Mixture} as:
\begin{align}
  & S[\{\rho\},\{\sfG\}] = \notag\\&\quad
  \max_{f\to\{\rho\},\{\sfG\}}\Big[
  -k_B \Trmix f \Big( \ln f 
  + \beta\sum_{\alpha=1}^M\sum_{i\in\Nalpha}\frac{\pv_i^2}{2m_\alpha}
  \Big)\Big].
\end{align}
It is again useful to subtract the ideal gas contributions and define
the excess entropy functional as
\begin{align}
  S_\rmexc[\{\rho\}, \{\sfG\}] &=
  S[\{\rho\}, \{\sfG\}] -
  \sum_{\alpha=1}^M S_\rmid[\rho_\alpha],
  \label{EQentropySexcMixture}
\end{align}
where $S_\rmid[\rho_\alpha]$ is the ideal gas entropy density
functional~\eqref{EQentropySidOneDefinition} depending on the partial
density distribution $\rho_\alpha(\rv)$. The `entropy of mixing' is
included in the present formulation by construction, cf.\ the additive
form of the partial ideal entropy contributions in the excess
defininition \eqref{EQentropySexcMixture}.

In generalization of the joint minimization
conditions~\eqref{EQentropyEL1rawAPX} and \eqref{EQentropyEL2rawAPX},
the present mixture formalism inplies the following species-resolved
versions:
\begin{align}
  \frac{\delta \Omega[\{\rho\}, \{\sfG\}]}{\delta \rho_\alpha(\rv)} 
  \Big|_{\{\rho_{\breve\alpha}\},\{\sfG\}} &= 0
  \quad {\rm (min)},
  \label{EQentropyEL1rhoGmixtureOmegaVersion}\\
  \frac{\delta \Omega[\{\rho\}, \{\sfG\}]}{\delta G_\aas(\rv)} 
  \Big|_{\{\rho\},\{\sfG_{\breve\alpha\breve\alpha'}\}} &= 0
  \label{EQentropyEL2rhoGmixtureOmegaVersion}
  \quad {\rm (min)}.
\end{align}
Per construction, at the minimum the value of grand potential is
recovered, $\Omega_0=\Omega[\{\rho\},\{\sfG\}]$ which generalizes the
one-component analog \eqref{EQOmegaFromEvaluationAtTheMinimizers}.

The relationships \eqref{EQentropyEL1rhoGmixtureOmegaVersion} and
\eqref{EQentropyEL2rhoGmixtureOmegaVersion} can be re-written, in
generalization of the joint Euler-Lagrange equations
\eqref{EQentropyEL1rhoGAPX} and \eqref{EQentropyEL2rhoGAPX}, in the
following more explicit forms:
\begin{align}
  c_\rho^{(\alpha)}(\rv;[\{\rho\}, \{\sfG\}]) &=
  \ln\big( \rho_\alpha(\rv)\Lambda_\alpha^d \big)
  + \beta V_{\rmext,\alpha}(\rv) - \beta\mu_\alpha,
  \label{EQentropyEL1rhoGmixture}\\
  c_G^{(\aas)}(r;[\{\rho\},\{\sfG\}]) &=
  \beta \phi_\aas(r),
  \label{EQentropyEL2rhoGmixture}
\end{align}
where the species-resolved thermal de Broglie wavelength is
$\Lambda_\alpha = h/\sqrt{2\pi m_\alpha k_BT}$ and we have defined
species-resolved one-body direct correlation functionals
$c_\rho^{(\alpha)}(\rv) = k_B^{-1}\delta
S_\rmexc/\delta\rho_\alpha(\rv)$ and $c_G^{(\aas)}(r) = k_B^{-1}
\delta S_\rmexc/\delta G_\aas(r)$, where we have suppressed the
functional dependencies on $\{\rho\}$ and $\{\sfG\}$ in the notation for
ease of reading.

For completness, species-resolved second-order direct entropy
correlation functions are obtained in generalization of the
one-component versions
\eqref{EQentropySecondOrderDirect1}--\eqref{EQentropySecondOrderDirect4}
via functional differentiation:
$c_{\rho\rho}^{(\alpha,\alpha')}(\rv,\rv') = \delta
c_\rho^{(\alpha)}(\rv)/\delta\rho_{\alpha'}(\rv)$,
$c_{\rho G}^{(\alpha,\alpha'\alpha'')}(\rv,r)=\delta
c_\rho^{(\alpha)}(\rv)/\delta G_{\alpha'\alpha''}(r)$,
$c_{G\rho}^{(\aas,\alpha'')}(r,\rv)=\delta
c_G^{(\aas)}(r)/\delta\rho_{\alpha''}(\rv)$, and
$c_{GG}^{(\aas,\alpha''\alpha''')}(r,r'')=\delta
c_G^{(\aas)}(r)/\delta G_{\alpha''\alpha'''}(r')$. These direct
correlation functionals are in particular relevant for the formulation
of the response and correlation theory, as described in
Sec.~\ref{SECresponseAndCorrelation} for pure systems, see in
particular the Ornstein-Zernike relations presented in
Sec.~\ref{SECOrnsteinZernikeExcessEntropy}.

\subsection{Quantum entropy functional}
\label{SECquantumEntropy}

We return to one-component systems and consider the statistical
mechanics of quantum $N$-body systems with interparticle interaction
potentials $u(\rv^N)$ that contain only pairwise contributions.  The
quantum description implies that the Hamiltonian
\eqref{EQentropyHamiltonianAPX} constitutes a Hilbert space
operator. The underlying fundamental position and momentum degrees of
freedom are operators that satisfy the canonical commutator
relationships: $[\rv_i, \rv_j]=[\pv_i,\pv_j]=0$ and
$[\rv_i,\pv_j]={\rm i}\hbar\delta_{ij}{\mathbf 1}$, where $[\hat A,
  \hat B]=\hat A \hat B - \hat B \hat A$ denotes the commutator of two
Hilbert space operators $\hat A$ and $\hat B$, ${\rm i}$ denotes the
imaginary unit, $\hbar$ indiates the reduced Planck constant,
$\delta_{ij}$ is the Kronecker symbol, and ${\mathbf 1}$ is the
$d\times d$ unit matrix.
The quantum trace in the grand ensemble is defined as
$\Trqm\,\cdot\,=\sum_{N=0}^\infty \sum_n \langle n | \,\cdot\,
|n\rangle/N!$, where $|n\rangle$ denotes a basis of the $N$-body
Hilbert space.
The quantum version of the Mermin functional $\Omega_M[f]$ is obtained
from the classical form~\eqref{EQentropyMerminFunctionalDefinitionAPX}
by replacement of the trace and of the Hamiltonian with their
respective quantum version.

The quantum version of the metadensity grand potential
functional~\eqref{EQentropyOmegaFunctionalFromMerminAPX} follows from
the joint constraints \eqref{EQentropyConstraint1APX} and
\eqref{EQentropyConstraint2APX}, where both the pair reduction
operator $\hat G(r)$ and the density operator $\hat\rho(\rv)$ retain
their respective formal definitions \eqref{EQentropyGhatAPX} and again
the trace is taken to be quantum mechanical. Also the grand potential
decomposition \eqref{EQentropyOmegaDecompositionAPX} and entropy
functional definition \eqref{EQentropyViaConstrainedMaximizationAPX}
constitute valid quantum versions.

Similarly, the quantum energy functional decomposition
\eqref{EQmeanEnergyFunctionalAPX}, total particle number functional
\eqref{EQmeanParticleNumberDensityFunctional}, and grand potential
form \eqref{EQmetaentropyOmegaViaEnergyEntropyAPX} continue to apply,
together with the metadensity minimization principle
\eqref{EQentropyMinimizationPrincipleAPX}, the pair of joint
Euler-Lagrange equations \eqref{EQentropyEL1rawAPX} and
\eqref{EQentropyEL2rawAPX}, the value of the grand potential at the
functional minimum~\eqref{EQOmegaFromEvaluationAtTheMinimizers}, and
the entropy splitting \eqref{EQentropySidSexcSplittingAPX} into ideal
and excess parts

An important difference between the quantum and classical descriptions
is that the classical ideal gas admits to obtain closed forms of
respective functionals for free energy, kinetic, and ideal entropy,
cf.\ Sec.~\ref{SECidealFunctionals}. This no longer holds in the
quantum case, as the ideal quantum gas is already non-trivial and
requires one typically to resort to the Kohn-Sham construction.

Retaining the definitions of the direct correlation functionals
\eqref{EQentropyDirectEntropic2APX} and
\eqref{EQentropyDirectEntropic2APX}, the Euler-Lagrange equations thus
change from their classical form \eqref{EQentropyEL1rhoGAPX} and
\eqref{EQentropyEL2rhoGAPX} to
\begin{align}
  c_\rho(\rv;[\rho,G])  &= 
  \frac{\delta \beta K[\rho]}{\delta \rho(\rv)}
  -\frac{\delta S_\rmid[\rho]/k_B}{\delta \rho(\rv)}
  + \beta V_\rmext(\rv) - \beta\mu,
  \label{EQentropyEL1rhoGAPXquantum}
\\
  c_G(r;[\rho,G]) &= \beta\phi(r).
  \label{EQentropyEL2rhoGAPXquantum}
\end{align}

The present formulation demonstrates structural consistency between
the quantum and classical versions of the entropy functional theory,
as one expects to find on general grounds. The parametric dependence
on temperature remains non-trivial for quantum systems, in contrast to
the classical case described in Sec.~\ref{SECidealFunctionals}.

}

\section{Proof by contradiction}
\label{SECentropyProofByContradiction}

We use the standard {\it reductio ad absurdum} argument
\cite{hansen2013, evans1979, mermin1965} to prove the relevant
functional map, as an alternative to the joint constrained search of
Secs.~\ref{SECoverview} and
\ref{SECentropyAppendixOmegaDerivation}. The aim is to show the
existence and uniqueness of the joint functional map $\rho(\rv),G(r)
\to V_\rmext(\rv),\phi(r)$. We follow Mermin \cite{mermin1965} and
Evans \cite{evans1979} and consider two different systems, an original
(unprimed) and a modified (primed) one. The unmodified system is as
before and the modified system is characterized by the external
potential $V_\rmext'(\rv)$ and the pair potential $\phi'(r)$. These
feature in the modified Hamiltonian $\Hams$, partition sum $\Xi'$,
grand potential $\Omega_0'$, and equilibrium phase space distribution
function $f_0'(\rv^N, \pv^N)$.

Mermin's functional \eqref{EQentropyMerminFunctionalDefinitionAPX}, when
expressed for this primed system, has the property
\begin{align}
  \Omega_0' = \Omega_M'[f_0'] < \Omega_M'[f_0],
  \label{EQentropyByContradictionOmegaZeroPrime}
\end{align}
where the strict inequality holds due to the assumption that the
primed and the unprimed systems are distinct.
The right hand side of the inequality
\eqref{EQentropyByContradictionOmegaZeroPrime} can be made more
explicit by writing out the Mermin functional
\eqref{EQentropyMerminFunctionalDefinitionAPX} as follows:
\begin{align}
  \beta \Omega_M'[f_0] 
  &= \Tr f_0 (\ln f_0 + \beta \Hams- \beta\mu N)\\
  &= \Tr f_0 \Big(\ln f_0 + \beta\sumi \frac{\pv_i^2}{2m}
  + \int_0^\infty dr \hat G(r) \beta\phi'(r)
  \notag\\&\qquad\qquad
  +\int d\rv \hat\rho(\rv) \beta V_\rmext'(\rv)\hat\rho(\rv)\Big)
  \label{EQentropyByContradictionIntermediate1}
  \\
  &= \Tr f_0 \Big(-\ln \Xi
  + \int_0^\infty dr \hat G(r) \beta\big(\phi'(r)-\phi(r)\big)
  \notag\\&\qquad\qquad
  +\int d\rv \hat\rho(\rv)
  \beta\big(V_\rmext'(\rv)-V_\rmext(\rv)\big)\hat\rho(\rv)\Big),
  \label{EQentropyByContradictionIntermediate2}
\end{align}
where we have first made the primed form of the Hamiltonian
\eqref{EQentropyHamiltonianRewrittenAPX} explicit in
\eqr{EQentropyByContradictionIntermediate1}.  Then using $\ln f_0 =
-\beta \Ham+\beta\mu N - \ln\Xi$ allows one to cancel the kinetic
energy and collect the interparticle and external potential
contributions in the two integrals in
\eqr{EQentropyByContradictionIntermediate2}.  Then observing that $\Tr
f_0 (-\ln \Xi) = \beta\Omega_0$ allows one to express the
inequality~\eqref{EQentropyByContradictionOmegaZeroPrime} as
\begin{align}
  \Omega_0' &< \Omega_0  + \int_0^\infty dr G(r)
  \big(\phi'(r)-\phi(r)\big)
  \notag\\&\qquad\;\;\,
  +\int d\rv \rho(\rv)\big(V'_\rmext(\rv) - V_\rmext(\rv)\big),
  \label{EQentropyByContradictionInequality}
\end{align}
where we recall the global pair reduction function $G(r) = \Tr f_0
\hat G(r)$ and the density profile $\rho(\rv) = \Tr f_0 \hat
\rho(\rv)$ being those of the original system in equilibrium.

We now make the assumption that these averages remain unchanged in the
primed system, $G(r) = \Tr f_0' \hat G(r)$ and $\rho(\rv) = \Tr f_0'
\hat \rho(\rv)$.  Then interchanging the primed and unprimed systems
in the inequality \eqref{EQentropyByContradictionInequality} gives
\begin{align}
  \Omega_0 &< \Omega_0'
  + \int_0^\infty dr G(r)  \big(\phi(r)-\phi'(r)\big)
  \notag\\&\qquad\;\;\,
  +\int d\rv \rho(\rv)\big(V_\rmext(\rv) - V'_\rmext(\rv)\big).
  \label{EQentropyByContradictionInequalityReversed}
\end{align}

The final step in the argument is to add the two inequalities
\eqref{EQentropyByContradictionInequality} and
\eqref{EQentropyByContradictionInequalityReversed}, which yields the
contradiction
\begin{align}
  \Omega_0 + \Omega_0' <   \Omega_0 + \Omega_0'.
\end{align}
It follows that the assumption that both $G(r)$ and $\rho(\rv)$ remain
unchanged in the primed system must have been false. This establishes,
for given $\beta,\mu, m$, the existence of the unique functional map:
\begin{align}
  \rho(\rv), G(r) \to V_\rmext(\rv), \phi(r).
\end{align}

From the standard argument \cite{evans1979, mermin1965} one can thus
conclude that thus the Hamiltonian is known and all equilibrium
properties of the system are determined.

\section{Conclusions}
\label{SECconclusions}

In conclusion, we have explored the consequences of the entropy-based
variational formulation of equilibrium statistical mechanics given in
Ref.~\cite{schmidt2026entropyShort}. The approach mirrors classical
density functional theory and it provides a description of the two
different (interparticle and external) potential energy contributions
to the grand potential on an equal footing. The mean interparticle
potential is incorporated into the variational framework in a formally
analogous way to the mean external potential energy. In contrast to
prior density functional explorations in the literature
\cite{percus1996, percus1994aspects, percus1989entropy}, we have shown
that this task does {\it not} require one to resolve explicitly the
inhomogeneous two-body density distribution $\rho_2(\rv,\rv')$. Often
$\rho_2(\rv,\rv')$ is viewed as the appropriate variational field that
is conjugate to the pair potential, expressed analogously as being a
function of two positions, $\phi_2(\rv,\rv')$.  We have here instead
made use of the fact that the pair potential $\phi(r)$ depends merely
on the scalar distance $r=|\rv-\rv'|$, such that one can identify
$\phi(r)=\phi_2(\rv,\rv')$ and make use of the fact that the conjugate
field to $\phi(r)$ is the global pair reduction function~$G(r)$
\cite{kampa2024meta, kampa2026pairmatching, kampa2026spherical}. This
vantage point implies a {\it significant} reduction of computational
complexity over any treatment that requires
resolving~$\rho_2(\rv,\rv')$ explicitly.

As a consequence of the generalization, besides the standard
functional dependence on the density profile~$\rho(\rv)$, the extended
variational `metadensity' functional dependence involves the
variational field $G(r)$, such that the grand potential $\Omega[\rho,
  G]$ acquires corresponding additional functional dependence. The
exact joint minimization principle
\eqref{EQentropyMinimizationPrincipleAPX} enables one to identify the
equilibrium state via determining both fields $\rho(\rv)$ and $G(r)$
that constitute the equilibrium solutions. Thereby the two conjugate
fields $V_\rmext(\rv)$ and $\phi(r)$ occur with mere linear functional
dependence in $\Omega[\rho,G]$, see
\eqr{EQentropyOmegaDecompositionAPX}.

We have laid out in detail the consequences of this variational
structure, including the splitting
\eqref{EQmetaentropyOmegaViaEnergyEntropyAPX} of the grand potential
into distinct entropy, energy, and chemical (particle number)
contributions \eqref{EQentropyViaConstrainedMaximizationAPX}--%
\eqref{EQmeanParticleNumberDensityFunctional}, which is structurally
equivalent to the standard thermodynamic sum. Due to the joint
minimization principle \eqref{EQentropyMinimizationPrincipleAPX}, the
theory features {\it two} coupled exact Euler-Lagrange equations
\eqref{EQentropyEL1rhoGAPX} and \eqref{EQentropyEL2rhoGAPX}, which
determine self-consistently the equilibrium forms of $\rho(\rv)$ and
$G(r)$ for given forms of the external potential $V_\rmext(\rv)$ and
the pair potential $\phi(r)$ at given thermodynamic conditions $\mu,
T$. Notably, the pair potential features only as an explicit
contribution in the `distance' Euler-Lagrange equation
\eqref{EQentropyEL1rhoGAPX}, which is formally analogous to the
occurrence of the external potential $V_\rmext(\rv)$ in the `position'
Euler-Lagrange equation \eqref{EQentropyEL2rhoGAPX}. The coupling
between the two Euler-Lagrange equations is via the two entropic
direct correlation functionals \eqref{EQentropyDirectEntropic1APX} and
\eqref{EQentropyDirectEntropic2APX} that follow systematically as
functional derivatives of the excess entropy functional
$S_\rmexc[\rho,G]$.

The present entropy functional framework not only determines the
thermodynamics via evaluating the grand potential functional at its
minimum \eqref{EQOmegaFromEvaluationAtTheMinimizers}, but it also
yields two- and higher-body correlation functions, as we have shown in
Sec.~\ref{SECsecondOrderResponseFunction} for the second order.  These
correlation functions are interrelatd via an entropic Ornstein-Zernike
structure, which consists of four coupled functional integral
equations, laid out in
Sec.~\ref{SECOrnsteinZernikeExcessEntropy}. Thereby the standard
correlation function of density fluctuations,
$H_2(\rv,\rv')=\cov(\hat\rho(\rv), \hat\rho(\rv'))$, features together
with generalized correlation functions that constitute covariances of
combinations of the local density operator~$\hat\rho(\rv)$ and the
global pair reduction operator~$\hat G(r)$, see
Eqs.~\eqref{EQentropyHrhorhoDefinition}--\eqref{EQentropyHGGdefinition}
for their specific definitions together with their role as response
functions that are generated by functional differentiation. The
closure of the entropic Ornstein-Zernike equations is provided by
second-order direct correlation functionals that follow as second
functional derivatives of the excess entropy functional
$S_\rmexc[\rho,G]$, see Sec.~\ref{SECsecondOrderDirect}.  Entropic
functional line integrals, as are presented in
Sec.~\ref{SECentropyFunctionalIntegration}, provide the inverse
operation that allows one to obtain the excess entropy from the
entropic direct correlation functionals. These functional integrals
are structurally no more involved than the standard free energy
functional line integrals \cite{evans1992, sammueller2023whyNeural},
which are well accessible in numerical work to obtain thermodynamic
quantities~\cite{sammueller2024pairmatching, sammueller2024attraction,
  robitschko2025mixShort, robitschko2026mixLong} and to target thermal
averages of general hyperobservables~\cite{sammueller2024hyperDFT,
  sammueller2024whyhyperDFT, sammueller2024multihyperDFT}.

We have given an equivalent reformulation of the entropy functional
based on a change of functional variables from the pair $\rho(\rv),
G(r)$ to the alternative $\rho(\rv), \go(r)$, where $\go(r)$ is a
global pair distribution function \eqref{EQentropySmallgAsRatio} that
is scaled with respect to the result for the ideal
gas~\eqref{EQentropyGiddy}.  We have developed in detail various
aspects of the theory, including formulations that include a localized
excess entropy density
(Sec.~\ref{SECentropyLocalizedEntrropyFunctional}), a reduced free
energy picture (Sec.~\ref{SECreducedFreeEnergyPicture}), the
description of the modifications that are required when the
interparticle potential features multi-body contributions
(Sec.~\ref{SECentropyManyBodyInteractions}-\ref{SECmanyBodyPracticalities}),
and the relationship with the recent spatially localized fluctuation
profiles (Sec.~\ref{SECentropyFluctuationProfiles}).  We have shown
that two of the arguably most iconic excess free energy density
functional approximations, the second-virial level low density
approximation (Sec.~\ref{SECpairEntropyApproximation}) and the
mean-field bilinear approximation (Sec.~\ref{SECmeanField}), follow
systematically from corresponding very simple excess entropy forms.

The relevant underlying functional map follows from a generalized
constrained search, where both the density distribution $\rho(\rv)$
and the global pair reduction function $G(r)$ constitute the
constraints under which the minimization is performed
(Sec.~\ref{SECentropyAppendixOmegaDerivation}). As an independent
alternative, we have provided an extended version of the Mermin-Evans
proof by contradiction \cite{evans1979, mermin1965, hansen2013} in
Sec.~\ref{SECentropyProofByContradiction}, which is based on the
inexistence of an alternative (primed) system that would differ in its
external potential $V_\rmext'(\rv)$ and pair potential $\phi'(r)$, yet
leads to identical $\rho(\rv)$ and~$G(r)$.

{\ms We have demonstrated the flexibility and generality of the
  entropy functional approach in Sec.~\ref{SECgeneralization} by
  treating: i)~many-body interparticle interactions via parametric
  dependence and a reduction operator formalism, ii) the equilibrium
  behaviour of general observables via formulation of an entropic
  hyperfunctional theory, iii) the equilibrium physics of classical
  mixtures using generalized, species-resolved variational fields, and
  iv) the thermal physics of many-body quantum systems. The present
  formal similarity between the classical and quantum versions of the
  entropy functional theory mirrors the situation of the recent
  statistical mechanical gauge invariance, where strong formal
  connections between the classical \cite{hermann2021noether,
    mueller2024gauge, mueller2024whygauge, nguyen2026,
    phamvan2026symmetry, phamvan2026active, maruyama2026} and quantum
  \cite{hermann2022quantum, mueller2025quantum,
    mueller2025quantumLong, phamvan2026quantumGeometry,
    phamvan2026exchangePrivateLindblad} physics are apparent.  In the
  present context, the analogy is important as it carries potential
  for investigating quantum-classical crossover from a new angle.

 }

As an outlook on possible future work, it would be interesting to
design machine learning schemes that aim to represent the entropic
functional relationships via neural functionals, i.e., neural networks
that accept functions as their input and yield the target output. Due
to the mathematical complexity of the entropic approach, including the
doubling of the Euler-Lagrange equation and the quadrupling of the
Ornstein-Zernike equations over the standard approach, this poses
significant challenges for the development of viable machine learning
schemes.  {\ms Thereby significant progress in the development of
  equivariant neural functionals is noteworthy
  \cite{cheng2026equivariant, weimar2026equivariant}.  The recent
  neural learning method for multivariate hyperdensity functionals
  demonstrates the feasibility and prowess of complex multi-stage
  training setups \cite{sammueller2024multihyperDFT}.}
Furthermore, investigating the consequences for solving inverse
problems \cite{henderson1974uniqueness, kampa2024meta,
  kampa2026pairmatching, kampa2026spherical}, soft matter design
tasks, {\ms and nonequilibrium properties \cite{schmidt2026power}} are
interesting points for future work.

A further inspiring thought is whether one could potentially be able
to carry out analytically the development of excess entropy
functionals.  Fundamental measure techniques offer an arguably very
flexible platform for addressing this task; see the excess free energy
functionals developed for penetrable spheres \cite{schmidt1999ps,
  rosenfeld2000ps}, for steep soft core repulsion
\cite{schmidt1999sfmt, schmidt2000sfmtMix, schmidt2000sfmtStructure},
and a variety of non-additive hard sphere mixtures
\cite{schmidt2000cip, schmidt2001wr, schmidt2004nahs, hopkins2010nahs,
  schmidt2011ternary}, as well as investigations of the geometrical
group structure that underlies fundamental-measure theory
\cite{schmidt2007peel, schmidt2011isometric}.  {\ms The mathematical
  structure of the exactly known free energy density functionals could
  provide further inspiration \cite{percus1976, robledo1981,
    vanderlick1989, percus1982, bakhti2012}.} We have restricted
ourselves to general equilibrium situations, addressing
non-equilibrium provides fertile ground \cite{schmidt2022rmp,
  diamant2026} as well as investigating responsive particles that can
adapt their properties according to external stimuli and specific
physical situations \cite{monchojorda2020, bley2021, bley2022,
  baul2021, lopezmolina2024, monchojorda2023}.  Also the concept of
excess entropy scaling \cite{rosenfeld1977scaling,
  rosenfeld1999scaling, chopra2010, pond2011, dyre2018scaling} and the
quasiuniversality of simple liquids \cite{dyre2016} might lend
themselves to be re-visited from the current angle. It is worth
exploring interconnections with the `shifting' gauge invariance of
statistical mechanics \cite{hermann2021noether, mueller2024gauge,
  mueller2024whygauge, hermann2022quantum, mueller2025quantum,
  mueller2025quantumLong, phamvan2026quantumGeometry, nguyen2026,
  phamvan2026symmetry, phamvan2026active, maruyama2026}. Recent exact
solutions for one-dimensional \cite{montero2019, montero2026dumbbell}
and quasi-one-dimensional \cite{montero2024,
  montero2024squarewellDisks, montero2024letter} systems could provide
further inspiration for making progress.
{\ms We leave important questions of representability, which have in
  particular been scrutinized in quantum density functional theory
  \cite{levy1979, ayers2005, ayers2007}, to future work. Also
  connections with the pair-level description \cite{ziesche1994,
    ziesche1996, ayers2005, nagy2006, higuchi2013, zhao2025} of
  quantum density functional theory are worth exploring.}

{\ms

Lastly, it would be very interesting to relate the present approach to
Rosenfeld's conjecture of the universality of the bridge
functional~\cite{rosenfeld1993bridge}, which was put forward in
classical work on phase separation of asymmetric binary hard spheres
\cite{rosenfeld1994prl, rosenfeld1995jcp}, strongly coupled plasmas
\cite{rosenfeld1996pre}, general liquid mixtures \cite{kahl1996pre},
the inverse problem of simple classical liquids
\cite{rosenfeld1997jpcm}, and in accuracy benchmarks against
simulation data~\cite{rosenfeld1998molPhys}.
We recall the bridge functional to consist of all beyond-second order
terms in the functional Taylor expansion of $F_\rmexc[\rho]$ around
the homogeneous bulk fluid.  The concept of the bridge functional was
shown to be relevant in a variety of contexts, from the construction
of reference functionals \cite{oettel2005} and weighted density
approximations \cite{yagi2021, yagi2022} to data-driven approaches
\cite{goodall2021closure}, the molecular density functional theory for
hydration and solvation properties \cite{borgis2020, borgis2021}, the
behaviour of patchy \cite{simon2024patchy} and penetrable particles
\cite{rosenfeld2000ps}, as well as the thermodynamics and liquid-gas
coexistence of water~\cite{zhao2013, jeanmairet2015}.}


\begin{acknowledgments}

I thank Bob Evans, Florian Samm\"uller, Stefanie M.\ Kampa, Johanna
M\"uller, Ana M. Montero, and Thomas Kriecherbauer for useful
discussions.  This work is supported by the DFG (Deutsche
Forschungsgemeinschaft) under Project No.~551294732.

\end{acknowledgments}


\end{document}